\documentclass{aa}

\usepackage{natbib}
\bibpunct{(}{)}{;}{a}{}{,} 

\usepackage{graphicx}
\usepackage{txfonts}
\usepackage{hyperref}
\hypersetup{
     colorlinks   = true,
     citecolor    = blue
}
\usepackage{units}
\usepackage{booktabs}

\begin{document}
\defcitealias{Schulreich:2017}{Paper~I}
\hyphenation{Schul-reich Breit-schwerdt Kor-schi-nek}

   \title{Numerical studies on the link between radioisotopic signatures on Earth and the formation of the Local Bubble}

   \subtitle{II.~Advanced modelling of interstellar \element[][26]{Al}, \element[][53]{Mn}, \element[][60]{Fe}, and \element[][244]{Pu} influxes as traces of past supernova activity in the solar neighbourhood}

   \author{M.~M.~Schulreich\inst{1} 
        \and
            J.~Feige\inst{1,2}
        \and 
            D.~Breitschwerdt\inst{1}
    }

   \institute{Zentrum f\"{u}r Astronomie und Astrophysik, Technische Universit\"{a}t Berlin, Hardenbergstra{\ss}e 36, 10623 Berlin, Germany\\
              \email{schulreich@astro.physik.tu-berlin.de}
         \and
             Museum f\"{u}r Naturkunde -- Leibniz-Institut f\"{u}r Evolutions- und Biodiversit\"{a}tsforschung, Invalidenstra{\ss}e 43, 10115 Berlin, Germany
    }

   \date{Received / Accepted}
   
   \titlerunning{Numerical studies on the link between radioisotopic signatures on Earth and the formation of the LB. II}
\authorrunning{M.~M.~Schulreich et al.}

 
  \abstract
   {Measurements of long-lived radioisotopes, which have grown rapidly in quantity and sensitivity over the last few years, provide a means, completely independent of other observational channels, to draw conclusions about near-Earth supernovae (SNe) and thus the origin of the Local Bubble (LB), our Galactic habitat. First and foremost in this context is \element[][60]{Fe}, which has already been detected across the Earth and on the Moon.}
   {The present study constitutes a significant step in further refining the coherent picture of the formation of the LB, constrained by radioisotopic anomalies, that we have drawn earlier, and is based on the most sophisticated initial conditions determined to date.}
   {Using \emph{Gaia} EDR3, we identified 14 SN explosions, with 13 occurring in Upper Centaurus-Lupus and Lower Centaurus-Crux, and one in V1062~Sco, all being subgroups of the Scorpius-Centaurus OB association. The timing of these explosions was obtained by us through interpolation of modern rotating stellar evolution tracks via the initial masses of the already exploded massive stars. We further developed a new Monte Carlo-type approach for deriving the trajectories of the SN progenitors, utilising a plethora of test-particle simulations in a realistic Milky Way potential and selecting explosion sites based on maximum values in six-dimensional phase-space probability distributions constructed from the simulations. We then performed high-resolution three-dimensional hydrodynamic simulations based on these initial conditions to explore the evolution of the LB in an inhomogeneous local interstellar medium and the transport of radioisotopes to Earth. The simulations include the effects of age- and initial mass-dependent stellar winds from the SN progenitors and additional radioisotopes (\element[][26]{Al}, \element[][53]{Mn}, and \element[][244]{Pu}) besides \element[][60]{Fe} using wind-derived and explosive yields from rotating models.}
   {From our modelling of the LB, we find as main results that (i) our simulations are consistent with measurements of \element[][60]{Fe}, in particular, a peak 2--\unit[3]{Myr} before present, as well as \element[][26]{Al}, \element[][53]{Mn}, and \element[][244]{Pu} data, (ii) stellar winds contribute to the distribution of radioisotopes and also to the dynamics of the LB, (iii) the solar system (SS) entered the LB about \unit[4.6]{Myr} ago, and (iv) the recent influx of \element[][60]{Fe}, discovered in Antarctic snow and deep-sea sediments, can be naturally explained by turbulent radioisotopic transport (in dust grains) originating mainly from the SN explosions and from the shock waves reflected at the LB shell.
   }
   {Our simulations not only support the recent hypothesis that the LB triggered star formation in the solar vicinity through its expansion, but also suggest that the second, separate \element[][60]{Fe} peak measured at 6--\unit[9]{Myr} ago was generated by the passage of the SS through a neighbouring superbubble (SB), possibly the Orion-Eridanus SB, prior to its current residence in the LB.}


   \maketitle
%

\section{Introduction}
\label{sec:intro}
The solar system (SS) currently resides within the so-called Local Bubble (LB), one of the numerous cavities in the interstellar medium (ISM) of our Milky Way, but also of other star-forming galaxies, that are filled with hot plasma and surrounded by a shell of cold, dusty gas. The existence of this superbubble (SB) was first proposed based on extinction mapping, which indicated that stars located within about \unit[100]{pc} of the Sun experience negligible reddening, suggesting very low local dust densities \citep[see the review by][and references therein]{Frisch:2011}. Measurements of the \ion{Na}{i} and \ion{Ca}{ii} absorption lines \citep{Welsh:2010} showed that the `Local Cavity' is also virtually free of neutral gas ($n < \unit[0.005]{cm^{-3}}$) out to 50--\unit[150]{pc} in the Galactic plane, where its dense outer `walls' rise. At high absolute Galactic latitudes, where the sampling is sparser, its extent is less well constrained, but is significantly more than in the plane, probably well over \unit[100]{pc}. Evidence for the high (million degree K) temperature in the LB interior came from the diffuse soft X-ray emission (energies $\sim$\unit[0.25]{keV} and higher) observed emanating from all directions of the sky \citep{Snowden:1997}. Since X-rays at these energies are immediately absorbed by the ions of for example C, O, N, and Fe (usually subsumed by their associated \ion{H}{i} column density), the detection of a substantial soft X-ray flux in the Galactic plane is an indication of the radiation's local origin. A conclusive explanation for all these features is that the LB was formed by stellar winds and supernovae (SNe) from nearby massive stars \citep[][and references therein]{Smith:2001}; for a review of LB observations and the role of the non-equilibrium ionisation (NEI) structure for cooling, see \cite{Breitschwerdt:2001} and \cite{Avillez:2012}.

Paradoxically, no cluster of massive stars could be found inside the LB. \cite{Fuchs:2006} therefore investigated the possibility whether the LB hosted one or more stellar moving groups in the past and found, through a kinematic analysis of the entire solar neighbourhood within a radius of \unit[200]{pc}, that young star clusters entered the present LB region approximately 10--\unit[15]{Myr} ago and that about 14--20 of their high-mass members have exploded since then, a result consistent with earlier studies by \cite{Maiz-Apellaniz:2001} and \cite{Berghoefer:2002}. The next major step was taken when \cite{Schulreich:2015}, \cite{Breitschwerdt:2016}, and \citet[hereafter \citetalias{Schulreich:2017}]{Schulreich:2017} succeeded in explaining the detection of live (undecayed) \element[][60]{Fe}, the most promising SN-produced radioisotope with half-life $\unit[2.61\pm 0.04]{Myr}$ \citep{Rugel:2009,Wallner:2015a}, on Earth within the framework of this model. The initial proposal to search for \element[][60]{Fe} and other long-lived (order of Myr) radioisotopes as tracers for near-Earth SN activity was put forth by \cite{Korschinek:1996} and \cite{Ellis:1996}.

To date, \element[][60]{Fe} has been found in deep-sea ferromanganese (FeMn) crusts, nodules, and sediments (including fossilised bacteria) from around the world \citep{Knie:1999,Knie:2004,Fitoussi:2008,Ludwig:2016,Wallner:2016,Wallner:2020,Wallner:2021}, in Antarctic snow \citep{Koll:2019}, in lunar soil \citep{Fimiani:2016}, in observations of the diffuse Galactic gamma-ray emission associated with its decay \citep{Wang:2007}, and in Galactic cosmic rays \citep[CRs;][]{Binns:2016}. The more recent and sensitive of these measurements in terrestrial archives showed not only the peak in \element[][60]{Fe} around 1.7--\unit[3.2]{Myr} ago, now known for over two decades, but also a second, separate peak around 6.5--\unit[8.7]{Myr} ago, and in addition low but significant \element[][60]{Fe} influx over the last \unit[33]{kyr}. Furthermore, two other radioisotopes of cosmic origin, \element[][53]{Mn} \citep[half-life $3.7\pm 0.4$\,Myr;][]{Honda:1971} and \element[][244]{Pu} \citep[half-life $81.3\pm 0.3$\,Myr;][]{Nesaraja:2017}, were detected in deep-sea material, with elevated concentrations of \element[][53]{Mn} in the same age range as the younger \element[][60]{Fe} signal \citep{Korschinek:2020}, whereas for \element[][244]{Pu} only the average influx over two rather broad time intervals of \unit[4.5]{Myr} each was determined for the last \unit[9]{Myr} \citep{Wallner:2021}, which, though overlapping with both \element[][60]{Fe} pulses, does not yet allow to make a statement about the exact timing of one or more potential \element[][244]{Pu} peaks.

The radioactive decay of live \element[][26]{Al} \citep[half-life $0.717\pm 0.024$\,Myr;][]{Basunia:2016} has been observed as broad angular distribution around the Galactic plane through its characteristic gamma-ray emission \citep{Plueschke:2001}. However,
its presence on Earth, which was expected to be contemporaneous with that of \element[][60]{Fe}, could not be confirmed \citep{Feige:2018}. This non-detection in combination with the existing \element[][60]{Fe} signal yielded a lower limit range of 0.10--0.33 for the $\element[][60]{Fe}/\element[][26]{Al}$ ratio produced in massive stars \citep{Feige:2018}, which is in agreement with the observed average Galactic $\element[][60]{Fe}/\element[][26]{Al}$ flux ratio of $0.184\pm 0.042$ \citep{Wang:2020}.

\element[][60]{Fe} is mainly synthesised during He and C shell burning and during the explosion of the star; both conditions providing high neutron densities ($>$$\unit[3\times 10^{10}]{cm^{-3}}$) required for the sequential neutron capture on stable \element[][58]{Fe} and on unstable \element[][59]{Fe} \citep{Limongi:2006}. \element[][26]{Al} is made in three different environments, that is during core H burning, H and C/Ne convective shell burning, and explosive Ne-burning, where the 
\element[][25]{Mg}(p,$\gamma$)\element[][26]{Al} reaction is primarily responsible for the \element[][26]{Al} production \citep{Limongi:2006, Diehl:2021}. Before being ejected in stellar explosions, \element[][26]{Al} is also released by stellar winds -- a process that becomes increasingly important with increasing initial mass \citep{Limongi:2006}. \element[][53]{Mn} is mainly produced in explosive Si and O burning as the SN shock wave moves through the inner parts of the star generating iron-group nuclei, of which \element[][53]{Fe} decays to \element[][53]{Mn} \citep{Meyer:2005}. The sources for the r-process isotope \element[][244]{Pu}, which is generated under explosive conditions, remain under debate. Recent studies point to rare events such as neutron star mergers (kilonovae (KNe)) or exotic types of SNe (e.g.~magnetorotational SNe) as dominant sites \citep{Hotokezaka:2015,Ji:2016,Cote:2021,Wallner:2015b,Wallner:2021}. 

While the cosmic influx of \element[][60]{Fe} onto Earth is dominated by the contribution of nearby SNe, the expected SN-derived influx of \element[][26]{Al} and \element[][53]{Mn} is overwhelmed by cosmogenic sources.  Spallation reactions between CRs and atmospheric Ar result in a continuous influx of \element[][26]{Al} \citep{Feige:2018}. Additionally, cosmogenic radioactive nuclei are produced in solid material in space, resulting in interplanetary dust-derived fluxes onto Earth of \element[][26]{Al} ($\sim$5 per cent of its atmospheric influx) and \element[][53]{Mn} \citep{Auer:2009, Korschinek:2020}. Anthropogenic \element[][244]{Pu} from recent nuclear weapon fallout has been detected in deep-sea FeMn crust layers with ages $>$\unit[1]{Myr} indicating downward migration \citep{Wallner:2021}. Such contamination could be avoided by complementary \element[][244]{Pu} measurements in lunar samples \citep{Wang:2021,Wang:2023}.

Earlier studies identified the stellar populations Upper Centaurus-Lupus (UCL) and Lower Centaurus-Crux (LCC) of the Scorpius-Centaurus (Sco-Cen) complex -- the closest OB association to the Sun \citep[100--200\,pc away;][and references therein]{Luhman:2022} -- as prime candidates for having hosted multiple sequential SNe that created the LB as well as the $\sim$3-Myr-old \element[][60]{Fe} signal on Earth (\citealt{Fuchs:2006}; \citealt{Schulreich:2015}; \citealt{Breitschwerdt:2016}; \citetalias{Schulreich:2017}). This finding was corroborated by the detection of a fast neutron star and its runaway companion that presumably originated from a stellar binary system of which the more massive component exploded about \unit[1.78]{Myr} ago at a distance of $\approx$\unit[107]{pc} within UCL \citep{Neuhaeuser:2020}. In addition, the nearby stellar association Tucana-Horologium (Tuc-Hor) was proposed to be a candidate for hosting a single SN event that created either the $\sim$3-Myr-old or the $\sim$7-Myr-old \element[][60]{Fe} signal \citep{Hyde:2018}.

In all cases, the SNe were expected to have occurred at moderate distances of 50--\unit[130]{pc} with progenitors exploding as electron-capture (EC) or core-collapse (CC) SNe (\citealt{Benitez:2002}; \citealt{Fry:2015}; \citealt{Schulreich:2015}; \citealt{Mamajek:2015}; \citealt{Breitschwerdt:2016}; \citetalias{Schulreich:2017}). At these distances, a SN blast wave can significantly alter the heliosphere upon impact and compress it up to $\sim$\unit[20]{au}, exposing parts of the outer SS directly to the SN plasma \citep{Miller:2022}. Hence, the SN material does not reach the Earth's orbit directly. Instead, \element[][60]{Fe}-bearing dust grains can decouple from the SN remnant and penetrate to the inner SS largely undeflected \citep{Fry:2016}. 

The recent influx of \element[][60]{Fe} has been proposed to originate from various sources. One hypothesis suggests that it arises from the passage of the SS through the Local Interstellar Cloud (LIC), assuming that the LIC is already enriched with \element[][60]{Fe} dust \citep{Koll:2019,Wallner:2020}. Another possibility is that it is due to dust-laden turbulent flows passing over the SS, excited by the SN shock waves themselves as well as their reflections off the outer shell of the LB, and permeating its cavity \citepalias{Schulreich:2017}.

The role played by the LB in the local ISM (LISM) recently received another facet, when \cite{Zucker:2022a} found that nearly all star-forming regions in the solar neighbourhood lie on the surface of the LB and that their young stars exhibit outward expansion mainly perpendicular to the LB's surface. These authors also presented calculations suggesting that the expansion of the LB is responsible for almost all of the nearby star formation.

In this paper, we present an updated model that builds upon our previous work, exploring the formation and evolution of the LB, as well as the associated transport of radioisotopes to Earth. We introduce novel implementations, including the consideration of stellar winds, additional radioisotopes besides \element[][60]{Fe}, and more. Our model now incorporates the latest stellar input data, such as new \emph{Gaia} astrometry, rotating stellar evolution tracks, and a modern value for the Sun's peculiar motion. Furthermore, we employ improved approaches for constraining the trajectories of the near-Earth SN progenitor stars. Our investigation specifically examines the transport through the LISM and deposition in deep-sea archives of four radioisotopes: \element[][26]{Al}, ejected from stellar winds and CC SNe; \element[][53]{Mn} and \element[][60]{Fe}, originating from CC SNe; and \element[][244]{Pu}, associated with rare events like KNe. Notably, we assume that \element[][244]{Pu} is present in the LISM at a constant concentration prior to the formation of the LB rather than being attributed to a specific event. This assumption is made due to the comparable half-life of \element[][244]{Pu} and the local interstellar mixing time, which is estimated to be on the order of \unit[100]{Myr} \citep{Avillez:2002}.

The paper is organised as follows. In Sect.~\ref{sec:model}, we give a detailed description of our updated LB model, including the Monte Carlo approach we apply for the first time to pinpoint the near-Earth SN explosion sites. The results obtained with the hydrodynamic simulations based on this framework are presented in Sect.~\ref{sec:results}. We conclude with a discussion of related issues in Sect.~\ref{sec:discussion} and a summary of our findings in Sect.~\ref{sec:summary}. 

\section{Model}
\label{sec:model}
\subsection{Unravelling the supernovae of the Local Bubble, again}
The basis of our study is a field star contamination cleaned sample of candidate members of the Sco-Cen complex selected by \cite{Luhman:2022} from the early installment of the third data release of \emph{Gaia} \citep[EDR3;][]{Gaia:2016,Gaia:2021}. To assign these stars to the individual Sco-Cen populations (Upper Scorpius (US), UCL/LCC, the V1062~Sco group, Ophiuchus, and Lupus), \citeauthor{Luhman:2022} looked for clustering in proper motion offsets and parallactic distance as well as celestial coordinates, and at the stars' positions in colour-magnitude diagrams. The resulting number of candidate members of each Sco-Cen population, $N_\mathrm{pop}$, is listed in Table \ref{tab:num}.
\begin{table*}
\caption{Properties of the stellar populations in the Sco-Cen complex. The table lists the population ages, the numbers of stars that satisfy the spatial, kinematic, and photometric criteria for membership in the populations as imposed by \cite{Luhman:2022} ($N_\mathrm{pop}$), the number of members with spectral types of M6 or earlier and initial masses above $\unit[0.5]{M_\sun}$ after accounting for incompleteness ($N_\mathrm{tot}$), the expected number of SNe ($N_\mathrm{SN}$), the current highest stellar initial mass in the samples ($M_\mathrm{max}$), and the IMF normalisation constant ($k$).}
\label{tab:num}      
\centering                        
\begin{tabular}{c c c c c c c}      
\hline\hline                
Population                 & Age\tablefootmark{a}  & $N_\mathrm{pop}$ & $N_\mathrm{tot}$ & $N_\mathrm{SN}$ & $M_\mathrm{max}$ & $k$\\
                           & (Myr)                 &                  &                  &                 & (M$_{\sun}$)     & (M$_{\sun}^{1.3}$)\\
\hline 
US                         & 11                    & 1147             & 401              & 0               & 5.7              & 221.1\\
Ophiuchus\tablefootmark{b} & 2--6\tablefootmark{c} & 160              & 67               & 0               & 1.9              &  43.0\\
Lupus                      & 6                     & 95               & 51               & 0               & 12.1             &  27.4\\
V1062~Sco                  & 20                    & 390              & 113              & 1               & 5.5              & 62.4\\
UCL/LCC                    & 20                    & 5273             & 1614             & 13              & 8.4              & 874.6\\ 
\hline
\end{tabular}
 \tablefoot{
   \tablefoottext{a}{As inferred by \cite{Esplin:2020}, \cite{Luhman:2020a}, and \cite{Luhman:2020} from comparing the Hertzsprung-Russell diagrams (HRDs) of low-mass stars in Sco-Cen and other nearby associations using \emph{Gaia} DR2.}
   \tablefoottext{b}{According to \cite{Luhman:2022}, the Ophiuchus sample is likely to be biased in favour of earlier spectral types, since those candidates feature the largest range of extinctions among the Sco-Cen populations and brighter, more massive stars can be detected at higher extinctions. To reduce such biases, we followed \citeauthor{Luhman:2022} and considered only candidates with extinctions less than $\unit[0.5]{mag}$ in the 2MASS $K_s$ band.}
   \tablefoottext{c}{For the calculations in this work, we used an average age of \unit[4]{Myr}.}
   }
\end{table*}
Following \citeauthor{Luhman:2022}'s assumption that each of these samples is 90 per cent complete for stars with estimated spectral types of M6 or earlier, and adding initial masses $\ge$$\unit[0.5]{M_\sun}$ as a further selection  criterion, we arrived at the total numbers, $N_\mathrm{tot}$, listed in Table \ref{tab:num}. The initial masses were determined such that they lie exactly on linearly interpolated versions of the \texttt{PARSEC}-\texttt{COLIBRI} stellar isochrones\footnote{\url{http://stev.oapd.inaf.it/cmd}} \citep{Bressan:2012,Chen:2014,Chen:2015,Tang:2014,Marigo:2017,Pastorelli:2019,Pastorelli:2020} calculated on the basis of the initial mass function (IMF) of \cite{Kroupa:2001}, for solar abundances at the ages of the stellar populations in question (see Table \ref{tab:num}). The absolute magnitudes required for this fitting were determined in the photometric system of the Two Micron All Sky Survey \citep[2MASS;][]{Skrutskie:2006}, since this is much less affected by extinction than that of \emph{Gaia}. We nevertheless still included the 2MASS $K_s$-band extinctions estimated by \cite{Luhman:2022} in the computation. In order to keep the samples as unbiased as possible by the catalogue cross-matching, only 2MASS point sources that are the closest match for the \emph{Gaia} sources within 3{\arcsec} were taken into account as a general rule. For parallactic distances, we adopted the geometric values estimated by \cite{Bailer-Jones:2021} from EDR3 parallaxes.

We next calibrated the IMF,
\begin{equation}
    \xi(m)\equiv\frac{\mathrm{d}N}{\mathrm{d}m}=k\,m^{-\alpha}\,,
    \label{eq:imfdef} 
\end{equation}
by integrating over the mass, $m$ (all masses are in units of solar masses), between 0.5 and $M_\mathrm{max}$, which is the current highest initial mass in each stellar population, as derived from the \cite{Luhman:2022} data (see Table \ref{tab:num}). Since the result ($N_\mathrm{tot}$) is already known, evaluating and rearranging the integral
\begin{equation}
    N_\mathrm{tot} = \int_{0.5}^{M_\mathrm{max}} k\,m^{-\alpha}\,\mathrm{d}m
\end{equation}
gives the desired normalisation constant
\begin{equation}
    k=\frac{(1-\alpha)\,N_\mathrm{tot}}{M_\mathrm{max}^{1-\alpha}-0.5^{1-\alpha}}\,,
\end{equation}
where $\alpha =2.3$ for the mass range under consideration \citep{Kroupa:2001}. 

The expected number of SNe, that is, the number of `missing' stars, is then calculated from
\begin{equation}
    N_\mathrm{SN} = \int_{M_\mathrm{min}}^{22} k\,m^{-\alpha}\,\mathrm{d}m\,,
\end{equation}
where $M_\mathrm{min} = \min(\max(9,M_\mathrm{max},M_\mathrm{ion}),22)$. The numerical factors that appear in the integration limits are based on recent parametric studies of CC models, according to which, in general, 95 per cent of Type II SNe should have initial masses in the range of 9--$\unit[22]{M_\sun}$\footnote{Above this initial mass range, there are only isolated, irregularly arranged intervals of variable width within which the stars explode as SNe. Outside these so-called islands of explodability, the stars collapse directly to black holes.} \citep[see][and references therein]{Straniero:2019}. $M_\mathrm{ion}$, on the other hand, denotes the minimum initial mass a SN progenitor star associated with a given population must have so that the explosion does not occur before less than $\unit[0.5]{Myr}$ and thus does not fall in a time range excluded by observed absorption features of the Li-like ions \ion{C}{IV}, \ion{N}{V}, and \ion{O}{VI}, as shown by \cite{Avillez:2012}. The resulting values of $N_\mathrm{SN}$ and $k$ are given for each population in Table \ref{tab:num}. 

Here we used the fact that initial masses dictate stellar lifetimes, which we estimated by linearly interpolating between the rotating Geneva stellar evolution models calculated by \cite{Ekstroem:2012}. Assuming that all stars in a population are born at the same time, the explosion time is simply the difference between the lifetime and the population age. Negative explosion times therefore refer to SN events in the past.

\begin{figure}
   \resizebox{\hsize}{!}
            {\includegraphics{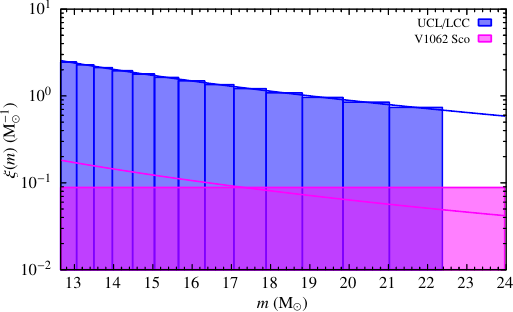}}
      \caption{IMFs of those Sco-Cen populations in which SN explosions should have already occurred (curves). Each mass bin of the respectively derived histograms (colours correspond) contains a single progenitor star.}
         \label{im:imf}
   \end{figure}

To establish the initial masses of the SN progenitor stars, we discretised the IMF using the first mean value theorem for definite integrals, according to which

\begin{equation}
    \Delta N=\int_{m_\mathrm{l}}^{m_\mathrm{r}} \xi(m)\,\mathrm{d}m=\xi(\bar{m})\,\Delta m\,,
    \label{eq:imfint}
\end{equation}
where $\Delta N$ is the number of stars contained in a mass bin of width $\Delta m=m_\mathrm{r} - m_\mathrm{l}>0$ and $\bar{m}$ is the mass for which the IMF achieves its mean value in the interval $(m_\mathrm{l},m_\mathrm{r})$. By substituting Eq.~(\ref{eq:imfdef}) into Eq.~(\ref{eq:imfint}) we obtain
\begin{align}
    \Delta N=k\,\int_{m_\mathrm{l}}^{m_\mathrm{r}} m^{-\alpha}\,\mathrm{d}m &=\frac{k}{1-\alpha}\,(m_\mathrm{r}^{1-\alpha}-m_\mathrm{l}^{1-\alpha})\nonumber\\&\overset{!}{=}k\,\bar{m}^{-\alpha}\,(m_\mathrm{r}-m_\mathrm{l})\,,
    \label{eq:deltaN}
\end{align}
which can be rearranged to yield
\begin{equation}
    \bar{m}=\left[(1-\alpha)\,\frac{m_\mathrm{r}-m_\mathrm{l}}{m_\mathrm{r}^{1-\alpha}-m_\mathrm{l}^{1-\alpha}}\right]^{1/\alpha}\,.
\end{equation}
Hence to quantify $\bar{m}$, we require the boundaries of the mass interval into  which it falls, which can be calculated using the iterative construction law also derivable from Eq.~(\ref{eq:deltaN}),
\begin{equation}
    m_\mathrm{r}=\left(m_\mathrm{l}^{1-\alpha}+\frac{1-\alpha}{k}\,\Delta N\right)^{1/(1-\alpha)}\,,
\end{equation}
where we start at the left boundary of the first interval, that is, at $m_\mathrm{l}=M_\mathrm{min}$, and the left boundary of the next interval corresponds precisely to the right boundary of the previous interval. By placing exactly one star in each mass bin (i.e.~$\Delta N=1$), we get the distribution with the highest probability \citep[see][]{Maiz-Apellaniz:2005}, with the values of $\bar{m}$ in the various bins serving as the initial masses, $M_\mathrm{ini}$, of the SN progenitor stars (see Fig.~\ref{im:imf}). These are listed in Table~\ref{tab:input} along with the explosion times and population affiliations of the perished stars. 

We emphasise that only the feedback of these 14 stars is included in the numerical simulations presented in Sect.~\ref{sec:results}. We may thereby miss the one or other star with initial mass above $\unit[22]{M_\sun}$ that collapses directly (i.e.~without exploding as a CC SN) into a black hole, blowing strong, fast stellar winds before doing so. However, as we show in Sect.~\ref{sec:results}, it is primarily the SNe that drive the evolution of the LB. And the comparatively short-lived \element[][26]{Al}, that would be released very early by these extremely massive stars, would have decayed too much by the time of the two \element[][60]{Fe} peaks to make any significant contribution to the terrestrial archives.

\begin{table*}
\caption{Input parameters for the LB simulations. The table lists the explosion times of the past SN events in the Sco-Cen complex ($t_\mathrm{exp}$), the Cartesian coordinates of their explosion centres with respect to the LSR ($X,Y,Z$), the distances of the explosion sites to the actual position of the Sun ($D$), the final peculiar velocity components of the SN progenitor stars ($U,V,W$), their initial masses ($M_\mathrm{ini}$), the \element[][26]{Al}, \element[][53]{Mn}, and \element[][60]{Fe} masses ejected during the explosions ($M_\mathrm{ej}$), the masses of the stellar remnants ($M_\mathrm{rem}$), and the stellar populations in which the explosions took place.}             
\label{tab:input}      
\centering                        
\begin{tabular}{c c c c c c c c c c c c c c}           
\hline\hline                
$t_\mathrm{exp}$ & $X$     & $Y$     & $Z$    & $D$     & $U$            & $V$            & $W$            & $M_\mathrm{ini}$ & \multicolumn{3}{c}{$M_\mathrm{ej}$}                                & $M_\mathrm{rem}$ & Pop.\tablefootmark{a}\\ 
(Myr)            & (pc)    & (pc)    & (pc)   & (pc)    & (km\,s$^{-1}$) & (km\,s$^{-1}$) & (km\,s$^{-1}$) & (M$_\sun$)       & \multicolumn{3}{c}{($\unit[10^{-4}]{M_\sun}$)}                     & (M$_\sun$)      &\\
\cmidrule{10-12}
                 &         &         &        &         &                &                &                &                  & $\element[][26]{Al}$ & $\element[][53]{Mn}$ & $\element[][60]{Fe}$ &\\
\hline 
$-10.16$         & $45.5$  & $27.3$  & $18.6$ & $236.4$ & $6.3$          & $-8.7$         & $6.6$          & $21.69$          & $1.00$               & $0.88$               & $2.84$               & $2.90$          & c\\
$-9.68$          & $64.5$  & $21.9$  & $9.8$  & $238.7$ & $3.1$          & $-7.4$         & $6.5$          & $20.42$          & $0.87$               & $0.83$               & $3.86$               & $2.92$          & c\\
$-8.90$          & $31.2$  & $-24.7$ & $16.1$ & $178.9$ & $-0.7$         & $-9.0$         & $0.4$          & $19.32$          & $0.75$               & $3.50$               & $3.72$               & $2.86$          & c\\
$-8.00$          & $30.3$  & $-26.5$ & $17.8$ & $161.2$ & $-0.5$         & $-9.3$         & $0.9$          & $18.34$          & $0.64$               & $7.35$               & $3.04$               & $2.75$          & c\\
$-7.20$          & $31.1$  & $-35.3$ & $15.6$ & $143.7$ & $-1.0$         & $-9.0$         & $0.5$          & $17.47$          & $0.55$               & $10.78$              & $2.43$               & $2.65$          & c\\
$-7.07$          & $113.1$ & $-15.6$ & $7.3$  & $220.9$ & $4.8$          & $-9.0$         & $4.5$          & $17.32$          & $0.54$               & $11.35$              & $2.33$               & $2.64$          & v\\
$-6.48$          & $33.5$  & $-40.5$ & $28.7$ & $136.4$ & $0.7$          & $-8.8$         & $3.4$          & $16.69$          & $0.47$               & $13.86$              & $1.88$               & $2.57$          & c\\
$-5.84$          & $31.2$  & $-50.8$ & $32.0$ & $123.0$ & $1.4$          & $-9.3$         & $2.5$          & $15.98$          & $0.39$               & $16.63$              & $1.39$               & $2.49$          & c\\
$-5.25$          & $33.5$  & $-57.8$ & $34.4$ & $115.8$ & $0.9$          & $-9.9$         & $3.2$          & $15.34$          & $0.32$               & $19.16$              & $0.94$               & $2.42$          & c\\
$-4.48$          & $37.5$  & $-57.6$ & $19.4$ & $101.2$ & $0.2$          & $-9.4$         & $0.3$          & $14.76$          & $0.28$               & $18.12$              & $0.71$               & $2.37$          & c\\
$-3.47$          & $40.9$  & $-65.9$ & $19.5$ & $92.2$  & $0.5$          & $-8.7$         & $0.4$          & $14.22$          & $0.25$               & $12.83$              & $0.74$               & $2.33$          & c\\
$-2.54$          & $33.5$  & $-74.3$ & $28.2$ & $83.1$  & $1.5$          & $-9.3$         & $0.9$          & $13.73$          & $0.23$               & $7.95$               & $0.77$               & $2.30$          & c\\
$-1.68$          & $99.3$  & $-18.0$ & $55.6$ & $128.5$ & $5.6$          & $-6.4$         & $4.5$          & $13.28$          & $0.21$               & $3.45$               & $0.79$               & $2.27$          & c\\
$-0.88$          & $47.7$  & $-87.1$ & $19.4$ & $96.0$  & $1.6$          & $-8.6$         & $-0.2$         & $12.85$          & $0.20$               & $0.72$               & $0.81$               & $2.25$          & c\\
\hline                                   
\end{tabular}
 \tablefoot{
   \tablefoottext{a}{c = UCL/LCC; v = V1062~Sco.}
   }

\end{table*}

\subsection{Constraining the paths of the near-Earth supernova progenitor stars using a Monte Carlo approach}
\label{sec:MC_approach}
   \begin{figure*}
   \resizebox{\hsize}{!}
            {\includegraphics{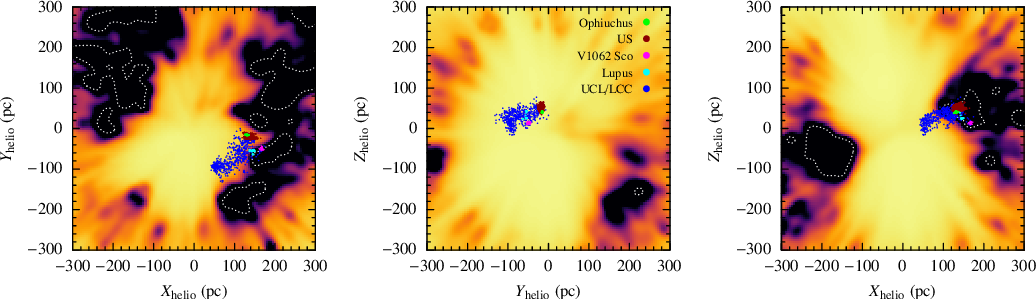}}
      \caption{Present-day heliocentric Galactic Cartesian coordinates of the field star purged member candidates of the Sco-Cen complex selected by \cite{Luhman:2022}, colour-coded by population. Plotted are only stars for which \emph{Gaia} measurements of radial velocities are available. The maps in the background show slices (from left to right along the plane through the Sun parallel to the Galactic mid-plane, the rotation plane, and the meridian plane) of the 3D dust density distribution in the LISM based on measurements of starlight absorption by dust \citep{Lallement:2019}. The colour-coding is black (light yellow) for high (low) dust densities. The dotted white contours correspond to a differential extinction of $\unit[0.003]{mag\,pc^{-1}}$ and delimit dense regions that are probably related to the LB's outer shell.}
         \label{im:obsmaps}
   \end{figure*}
   
It remained to determine the trajectories of the SN progenitors. To do so, we used the candidate members of the populations in Sco-Cen selected by \cite{Luhman:2022} that have measured radial velocities. It was demonstrated by the same author that these stars (469 in US, 96 in Ophiuchus, 52 in Lupus, 27 in V1062~Sco, and 542 in UCL/LCC) provide a good sampling of the population to which they belong. We followed the procedure described in Sect.~4.1.7 of \cite{Hobbs:2021} to calculate for each star from its astrometric parameters (incl.~radial velocity) and their observational uncertainties the rectangular space coordinates and velocity components together with their errors. Assuming these errors to be normal-distributed, which should be approximately the case for the nearby stars considered here \citep[see][]{Luri:2018}, we generated 10\,000 random realisations of each population using the Marsaglia polar method \citep{Marsaglia:1964}. The mean current positions of the stars are shown in Fig.~\ref{im:obsmaps} in projection onto cuts through the three-dimensional (3D) local interstellar dust density distribution as derived by \cite{Lallement:2019}.

We used the leapfrog integrator of the galactic-dynamics \texttt{python} package \texttt{galpy}\footnote{\url{https://github.com/jobovy/galpy}} \citep{Bovy:2015} to numerically trace each of these realisations separately back in time to the birth time of the respective stellar population. Since the quality of this kind of test-particle simulations depends crucially on the choice of a realistic mass model for the Galactic disc, we did not use any of the `standard' \texttt{galpy} potentials but implemented the model (without ring density structure) of \cite{Barros:2016}, which was developed specifically for this purpose. Its main idea is to construct the Milky Way disc by a superposition of distinct models of Miyamoto-Nagai (MN) discs \citep{Miyamoto:1975} fitted to observations; `these models are comprised of combinations of three MN-discs for each one of the four subcomponents: thin, thick, \ion{H}{I}, and H$_2$ discs yielding 12 MN-discs in total' \citep{Barros:2016}. The bulge and dark halo components, on the other hand, are classically described by a Hernquist potential \citep{Hernquist:1990} and a spherical logarithmic potential \citep[e.g.][]{Binney:2008}, respectively. 

Since interstellar gas has usually only small peculiar motions, the LISM -- and with it the LB -- will basically co-rotate with the local standard of rest (LSR) around the Galactic centre \citep[GC;][]{Fuchs:2006}. This is why we performed all our calculations in the LSR frame. Its origin is in the Galactic mid-plane below the current position of the Sun, which lies $Z_\sun\approx\unit[20.8]{pc}$ \citep{Bennett:2019} above. For the Sun, whose motion we also tracked, we adopted a Galactocentric distance of $R_0=\unit[8]{kpc}$ \citep{Malkin:2013} and a peculiar velocity relative to the LSR of $\vec{V}_\sun=(U_\sun,V_\sun,W_\sun)=\unit[(11.10,12.24,7.25)]{km\,s^{-1}}$ \citep{Schoenrich:2010} for the present time ($t=0$); we used the usual right-handed coordinate system, in which the $X$-axis points towards the GC, the $Y$-axis into the direction of Galactic rotation, and the $Z$-axis towards the North Galactic Pole. For the rotation velocity of the LSR we followed \cite{Barros:2016} and adopted a value of $V_0=\Omega_0\,R_0=\unit[230]{km\,s^{-1}}$, which is based on a local angular rotation velocity of $\Omega_0=\unit[28.7]{km\,s^{-1}\,kpc^{-1}}$, as estimated from direct measurements of the proper motion of Sgr~A$^{*}$ \citep{Reid:2004}.

\begin{figure*}
   \resizebox{\hsize}{!}
            {\includegraphics{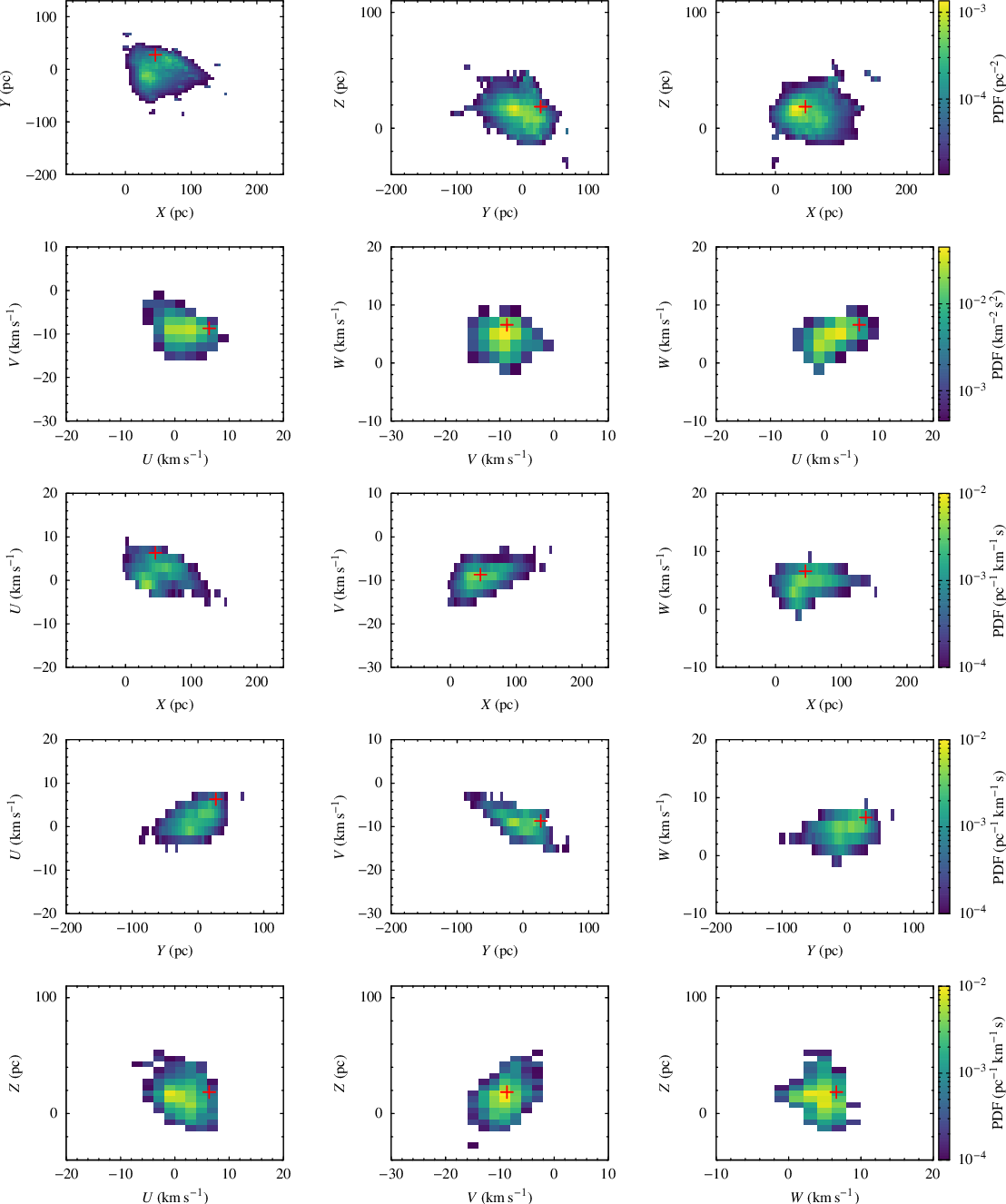}}
      \caption{Marginal phase-space PDFs for UCL/LCC at $t=\unit[-10.16]{Myr}$, computed by tracing back in time 10\,000 realisations of the population. The colour bar is the same for all plots in a row. The red crosses mark projections of the bin where the 6D phase-space PDF reaches its maximum value ($\sim$$\unit[1.9\times 10^{-6}]{pc^{-3}\,km^{-3}\,s^3}$), which arises from 10\,040 of the total 5\,420\,000 pseudostars ($\sim$0.2 per cent). From a purely stellar-statistical point of view, a SN at this particular time should have originated from a massive star at approximately this location with approximately this peculiar velocity.}
         \label{im:pspdf}
   \end{figure*}

At the time of explosion of each SN, we paused all back-calculations of the corresponding population and computed the six-dimensional (6D) phase-space probability density function (PDF) for the population's combined realisations. As the explosion site and final velocity of the SN progenitor star we took the point where the PDF has its maximum. Since this is strictly speaking not a point but an extended region of, in our case, $(\unit[5]{pc})^3\times(\unit[2]{km\,s^{-1}})^3$ size, we determined the six coordinate values by a random draw from this particular phase-space volume element. For the sake of demonstration, we show in Fig.~\ref{im:pspdf} projections of the phase-space PDF for UCL/LCC onto two-dimensional (2D) planes (i.e.~marginal PDFs), generated for pinning down a putative SN event at $t=\unit[-10.16]{Myr}$. Red crosses mark the most probable phase-space coordinates, which are listed together with those of all other possible past SNe in the Sco-Cen complex in Table \ref{tab:input}.

Finally, by calculating each of these missing, hypothetical stars back to the birth time of the population to which they belong, we obtained the initial values for their motion. Owing to the time-reversibility and symplectic nature of leapfrog integration, we could be sure to pass through the exact explosion sites again when calculating these stars forward in time.

We note that the diffuseness of the PDFs shown in Fig.~\ref{im:pspdf} is related to the fact that the further back (or ahead) in time a stellar population is calculated, the more the trajectories of its members are affected by the observational errors inherent in the initial conditions. This effect, which leads to an artificial broadening of the population in phase space, can already be observed in PDFs that are generated from a single realisation of the population, such as the one given by the mean phase-space coordinates. If one were to limit oneself to this, however, one would discard additional information given by the observational errors and thus deprive oneself of the possibility of reconstructing the conditions at the time of birth of the population and thus its entire dynamic evolution more precisely (whereby the accuracy increases with the number of realisations used). Needless to say that the closer one gets to the present time, the less diffuse the phase-space PDFs generally become.

\subsection{Hydrodynamic simulation setup}
As in \citetalias{Schulreich:2017}, we performed 3D hydrodynamic simulations of the formation of the LB using the Eulerian tree-based adaptive mesh refinement code \texttt{RAMSES}\footnote{\url{https://bitbucket.org/rteyssie/ramses}} \citep{Teyssier:2002}. The gas, which was assumed to obey a perfect equation of state with polytropic index $\gamma=5/3$, was evolved with a second-order unsplit Godunov scheme \citep{Leer:1979}. The latter was coupled with an approximate Riemann solver that is less diffusive than the one applied in \citetalias{Schulreich:2017} \citep[Harten-Lax-van Leer-Contact, or HLLC for short, instead of Local Lax-Friedrichs;][]{Toro:1994}, using a MinMod total variation diminishing scheme \citep{Roe:1986} to interpolate the cell-centred quantities at the cell interfaces. The time steps are restricted by the Courant-Friedrichs-Lewy condition (we set the Courant factor to 0.8) and were enforced by us to be synchronous (i.e.~the coarser level is updated using the same time step as the finer level). Apart from switching the solver, we also extensively revised and improved the numerical model, as described below.

\subsubsection{Background medium}
For the background medium, we chose a compromise between a homogeneous environment and one that has evolved self-consistently until convergence as a result of SN-driven turbulence. The latter is not only extremely computationally intensive to produce, but also poses the challenge of selecting a region where the evolution of the LB is not overly influenced by surrounding gas flows, which, although representative of the LISM dynamics in general, can be an obstacle in reproducing a specific scenario. The main reason behind this is the inherent inconsistency of the LB originating in a region whose properties do not reflect the true history of the LISM, but are chosen ad hoc for solely being an inhomogeneous environment in a statistical sense.

We therefore used a smoothly stratified medium that, under the assumption of isothermality and an ideal gas, is in hydrostatic equilibrium with the total gravitational potential, $\Phi$, of \cite{Barros:2016} outlined in Sect.~\ref{sec:MC_approach}, whose negative gradient is added to the Euler equations as an external force term (to maintain stability, we ignored the self-gravity of the gas). Since we limited our simulations to a cube-shaped cutout of the Galactic disc, whose edge length of \unit[800]{pc} is small compared to the total radius of the disc (with the lateral extent of the LB being even smaller), we neglected any radial gradients and considered only vertical ($z$-dependent) distributions valid at the solar circle (i.e.~at the Galactocentric radius $r=R_0$). This leads to the initial mass density and pressure profiles
\begin{equation}
	\rho(z)=\rho_0\,\exp\left[\frac{\Phi_0-\Phi(z)}{\mathcal{R}\,T_\mathrm{ini}}\right]	
\end{equation}
and
\begin{equation}
	P(z)=\mathcal{R}\,T_\mathrm{ini}\,\rho(z)	\,,
\end{equation}
respectively, where $\mathcal{R}\equiv k_\mathrm{B}/(\mu\,m_\mathrm{H})$ denotes the specific gas constant ($k_\mathrm{B}$ is the Boltzmann constant, $\mu$ is the mean molecular weight, and $m_\mathrm{H}$ is the mass of a hydrogen atom) and the subscript zero refers to mid-plane values. For the initial temperature, $T_\mathrm{ini}$, we adopted \unit[8000]{K}, which is a typical value for the warm neutral medium.

\subsubsection{Stellar winds}
All stars (including the Sun) were treated as collisionless particles in the simulations and moved along the trajectories already determined with \texttt{galpy}, since these are based on the full, cylindrically symmetric Milky Way potential of \cite{Barros:2016}. To set up the energetic winds blown by the massive stars, at least two parameters are required: the mass-loss rate from the stellar atmosphere, $\dot{M}$, and the velocity of the wind at infinity (terminal velocity), $\varv_\infty$, with both parameters depending not only on the initial masses of the stars but also on their individual evolutionary stages. 

The mass-loss rate is already given accordingly tabulated in the stellar tracks of \cite{Ekstroem:2012}. The same applies to the surface abundance of the radioisotope \element[][26]{Al}. For setting the terminal velocity, we used the wind prescription \texttt{wind08} of \cite{Voss:2009}, which first, following \cite{Leitherer:1999}, roughly classifies stars into luminous blue variables (LBVs) and Wolf-Rayet (WR) stars based on their mass-loss rate and effective temperature. The WR stars are then, following \cite{Smith:1991} and \cite{Leitherer:1999}, further divided into subclasses based on their surface abundances of H, C, N, and He, with the WR-wind velocities estimated by \cite{Niedzielski:2002}. Stars that do not fall into these categories are divided into hot and cool stars based on their effective temperature \citep{Lamers:1995}, with their terminal velocity depending on the escape velocity at their surfaces \citep[for details see][]{Voss:2009}. We calculated the surface escape velocities using the values of effective temperature ($T_\mathrm{eff}$), mass ($M$), and luminosity ($L$), tabulated for the individual stellar ages in the evolutionary tracks of \cite{Ekstroem:2012} via
\begin{equation}
	\varv_\mathrm{esc}=\left(\frac{2\,G\,M}{R}\right)^{1/2}\overset{L=4\pi R^2 \sigma_\mathrm{SB} T_\mathrm{eff}^4}{=}2\,T_\mathrm{eff}\,(G\,M)^{1/2}\,\left(\frac{\pi\,\sigma_\mathrm{SB}}{L}\right)^{1/4}\,,
\end{equation}
where $G$ is the gravitational constant, $R$ is the stellar radius, and $\sigma_\mathrm{SB}$ is the Stefan-Boltzmann constant. With the mass-loss rate, terminal velocity, and \element[][26]{Al} surface abundance evaluated for the initial mass and stellar age values listed in the \cite{Ekstroem:2012} data, we computed the amounts of mass (total and radioisotopic), momentum, and energy injected by each wind-blowing star per time step into its enclosing grid cell on the fly during the simulation using bilinear interpolation between data points\footnote{This approach is basically the same as that chosen by \cite{Fierlinger:2014} for simulating a different SB.}.

\subsubsection{Supernova explosions}
In our model, once the lifetime of a massive star has expired, it explodes as an SN, releasing the canonical explosion energy of $E_\mathrm{SN}=\unit[10^{51}]{erg}$ into a single grid cell in pure thermal form. The total mass ejected in the SN was taken to be the difference between the total mass at the last point of the stellar evolution model and the mass of the remnant, which depends on the initial mass of the progenitor star and is based on the rotating models of \citeauthor{Limongi:2018} (\citeyear{Limongi:2018}; their recommended set R, together with with their intermediate value for the initial equatorial rotation velocity of $\unit[150]{\,km\,s^{-1}}$). The same applies to the explosive yields of \element[][26]{Al}, \element[][53]{Mn}, and \element[][60]{Fe} (see Table \ref{tab:input}). A constant extrapolation was used for initial masses below $\unit[13]{M_\sun}$ for which no data are available. However, this was only necessary for the last SN explosion in our model, whose progenitor star has an initial mass of $\unit[12.85]{M_\sun}$ (see Table \ref{tab:input}).

\subsubsection{Radiative cooling}
To treat the radiative cooling of the gas in a more standardised way, we switched to the publicly available cooling library \texttt{GRACKLE}\footnote{\url{https://github.com/grackle-project/grackle}} \citep{Smith:2017}. We adopted the equilibrium cooling mode of the library (metals are assumed to be in ionisation equilibrium), which is based on cooling rates precomputed using the \texttt{Cloudy} photo-ionisation code \citep{Ferland:2013}. Cooling rates are tabulated as a function of gas density, temperature, and metallicity. They include contributions from both primordial cooling and metal line cooling, which is determined for solar abundances. The latter contribution can be scaled to the gas metallicity, which, however, we left as solar everywhere in our simulations. By prohibiting the gas from cooling down further than $T_\mathrm{init}$ we ensured that the background medium remains in hydrostatic as well as thermal equilibrium.

\subsubsection{Radioactive isotopes}
\label{sec:radioiso}
As in \citetalias{Schulreich:2017}, we treated the radioisotopes, which in reality must be adsorbed to dust grains to enter the heliosphere \citep{Fields:2008}, in our simulations as passive scalars -- contaminants at such low concentrations that they do not affect the flow but are carried by the fluid according to an advection-diffusion equation. There was a claim in the recent literature (\citealt{Fry:2020}; see also Sect.~\ref{sec:compmod}) that the perfect coupling between SN dust and plasma implicitly assumed by this would not be appropriate for setups such as ours. We disagree with this statement in Appendix \ref{app:A}.

To isolate the radioisotopic contributions of the stellar winds and SNe that formed the LB, we set the initial concentrations of \element[][26]{Al}, \element[][53]{Mn}, and \element[][60]{Fe} in the background medium to zero. For \element[][244]{Pu}, which was presumably released by a KN prior to the formation of the LB and thus was already present in its region of influence in an unknown amount (referred to by \citealt{Wang:2021} as the `two-step scenario'), we first assumed an initial concentration of 100 per cent. After running the simulations, we then scaled the concentration so that the ratio of the local interstellar fluxes of \element[][244]{Pu} and \element[][60]{Fe} integrated over the age interval of $0$--$\unit[4.57]{Myr}$ exactly matches the value of about $5.3\times 10^{-5}$ measured by \cite{Wallner:2021} for the same time span. This allows us not only to track the turbulent mixing of \element[][244]{Pu} in absolute numbers but also to draw conclusions about its initial concentrations.

Since we explicitly included the motion of the Sun relative to the LSR, unlike \citetalias{Schulreich:2017}, the flux of radioisotopes now had to be measured not at a fixed point in the computational domain but always in the grid cell where the Sun is currently located\footnote{This is analogous to what \cite{Chaikin:2022} considered in a smoothed-particle hydrodynamics (SPH) framework (see Sect.~\ref{sec:compmod}).}. The flux of a species $i$ at the heliosphere boundary (local interstellar flux) is thus calculated as
\begin{equation}
	F_i=\frac{\rho\,\tilde{u}\,C_i}{A_i\,m_\mathrm{u}}\,.
\end{equation}
Here, $A_i$ is the atomic number of the species, $m_\mathrm{u}$ is the atomic mass unit, and $\tilde{u}\equiv|\vec{u}-\vec{V}_\sun|$ is the flow speed in the rest frame of the Sun. The variables $\rho$, $\vec{u}$, and $C_i$ denote the (total) gas mass density, flow velocity, and species concentration, respectively, all evaluated in the grid cell enclosing the Sun. It is important to note that we made sure that the Sun, as well as all other considered stars, always reside in grid cells at the highest refinement level. In particular, we enforced this for all grid cells around a particle within a radius of $\unit[10]{pc}$, which guarantees maximum accuracy both in the determination of the radioisotopic fluxes and in the initialisation of the stellar feedback processes. Outside these spherical regions, the resolution of the flow is dynamically controlled by criteria based on the steepness of both density and pressure gradients. We also note that the value of $F_i$ already includes the radioactive decay that the species has experienced since its release in the ISM, as this process is directly accounted for in the simulations.

After penetration of the heliosphere, (isotropically assumed) fallout onto the Earth, and incorporation into a natural reservoir such as a deep-sea crust or sediment, the current measurable column density of $i$ in a layer $\ell$ covering the age range $\Delta a_\ell=(a_{\ell,\mathrm{l}},a_{\ell,\mathrm{u}})$ is
\begin{equation}
	N_i(\Delta a_\ell)=\frac{f_i}{4}\,\int_{a_{\ell,\mathrm{l}}}^{a_{\ell,\mathrm{u}}}F_i(a)\,\exp(-\lambda_i\,a)\,\mathrm{d}a\,,
	\label{eq:species_colden}
\end{equation}
where the factor $1/4$ is derived from the ratio of the cross-sectional area of the Earth to its surface and $0<f_i<1$ is the fraction of $i$ that overcomes all these filtering processes (`survival fraction'). The exponential function in the integrand of Eq.~(\ref{eq:species_colden}) accounts for the radioactive decay the species has undergone since its deposition on Earth ($\lambda_i \equiv \ln(2)/t_{1/2,i}$, with $t_{1/2,i}$ denoting the species' half-life). Division of Eq.~(\ref{eq:species_colden}) by the physical thickness of the associated layer, $\Delta d_\ell$, gives the present-day distribution of the number density of $i$ throughout the reservoir,
\begin{equation}
	n_i(\Delta a_\ell)=\frac{N_i(\Delta a_\ell)}{\Delta d_\ell}\,.
\end{equation}
This is often given in relation to the mean number density of the corresponding stable element $j$, 
\begin{equation}
n_j=\frac{w_j\,\rho_\mathrm{r}\,N_\mathrm{A}}{m_j}\,,
\end{equation}
where $w_j$ is the weight fraction of $j$, $\rho_\mathrm{r}$ is the mean density of the reservoir, $N_\mathrm{A}$ is the Avogadro constant, and $m_j$ is the molar mass of $j$.

\section{Results}
\label{sec:results}
We conducted a series of simulations with a theoretical spatial resolution of $\unit[0.78]{pc}$, in which we varied the atomic hydrogen number density in the Galactic mid-plane between $0.1$ and $\unit[1]{cm^{-3}}$ in steps of $\unit[0.1]{cm^{-3}}$ and found that for $n_\mathrm{H,0}=\unit[0.7]{cm^{-3}}$ the calculations best reproduce both the observable properties of the LB and the radioisotopic measurements available. Simulations with a resolution twice as high did not lead to significant changes in the physical quantities of interest. Therefore, we focus on this best-fitting run for the rest of this paper.

\subsection{Evolution and properties of the Local Bubble}
   \begin{figure*}
   \centering
   \resizebox{0.96\hsize}{!}
            {\includegraphics{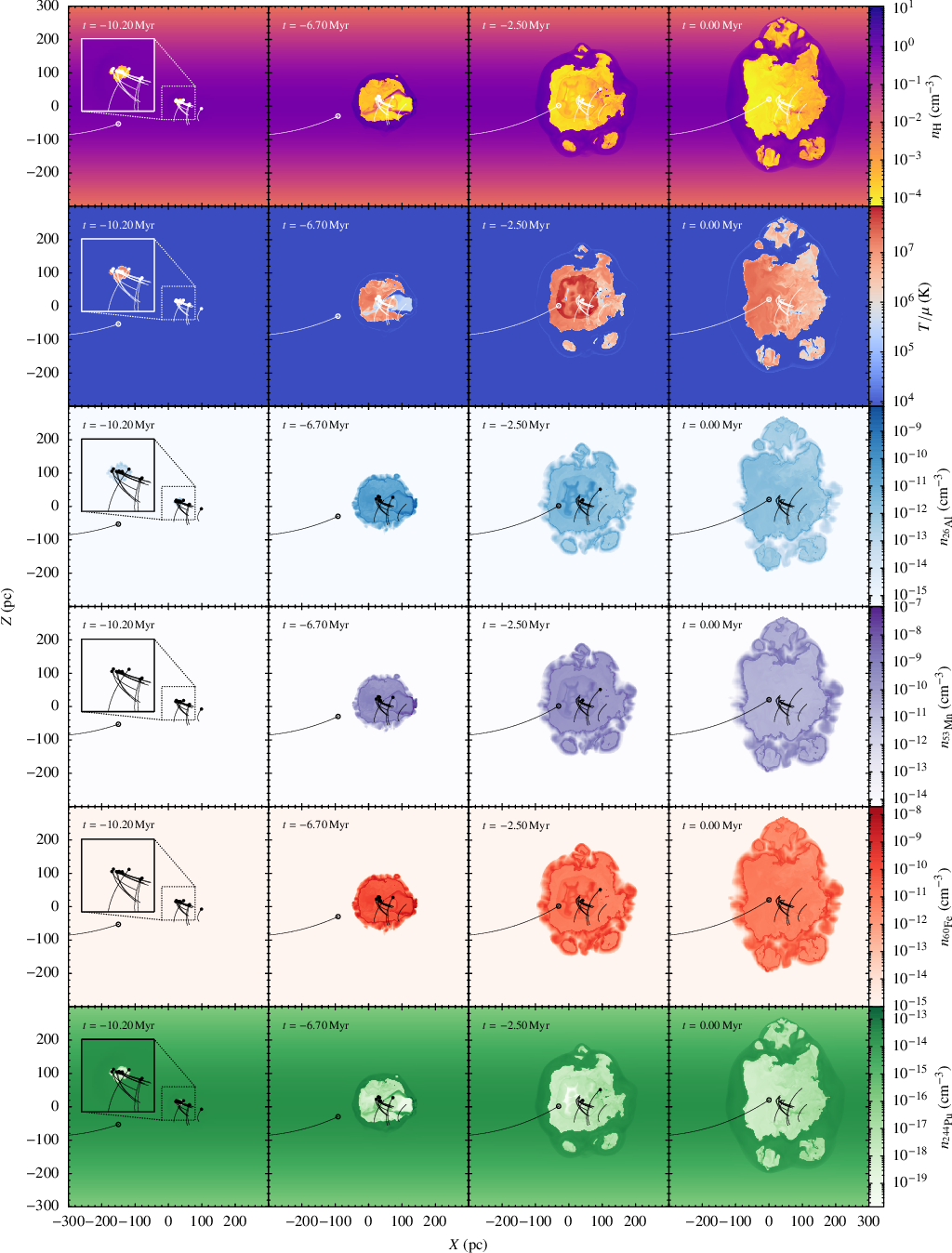}}
      \caption{2D axis-aligned slices at $Y=0$ through the 3D computational domain showing colour-coded (\emph{from top to bottom}) the atomic hydrogen number density, temperature, and number densities of the radioisotopes \element[][26]{Al}, \element[][53]{Mn}, \element[][60]{Fe}, and \element[][244]{Pu} in the LB region at different times in the past, as extracted from our fiducial simulation. The overplotted symbols represent for the respective time the projected positions of the Sun (empty circle) and of the SN progenitor stars that have not exploded by then (filled circles). Projected trajectories are shown as thin solid lines. The inlay in the first-column panels shows a magnification of the boxed area with side length 100\,pc.
      }
         \label{im:ts-sy}
   \end{figure*}
  
   \begin{figure*}
   \centering
   \resizebox{0.96\hsize}{!}
            {\includegraphics{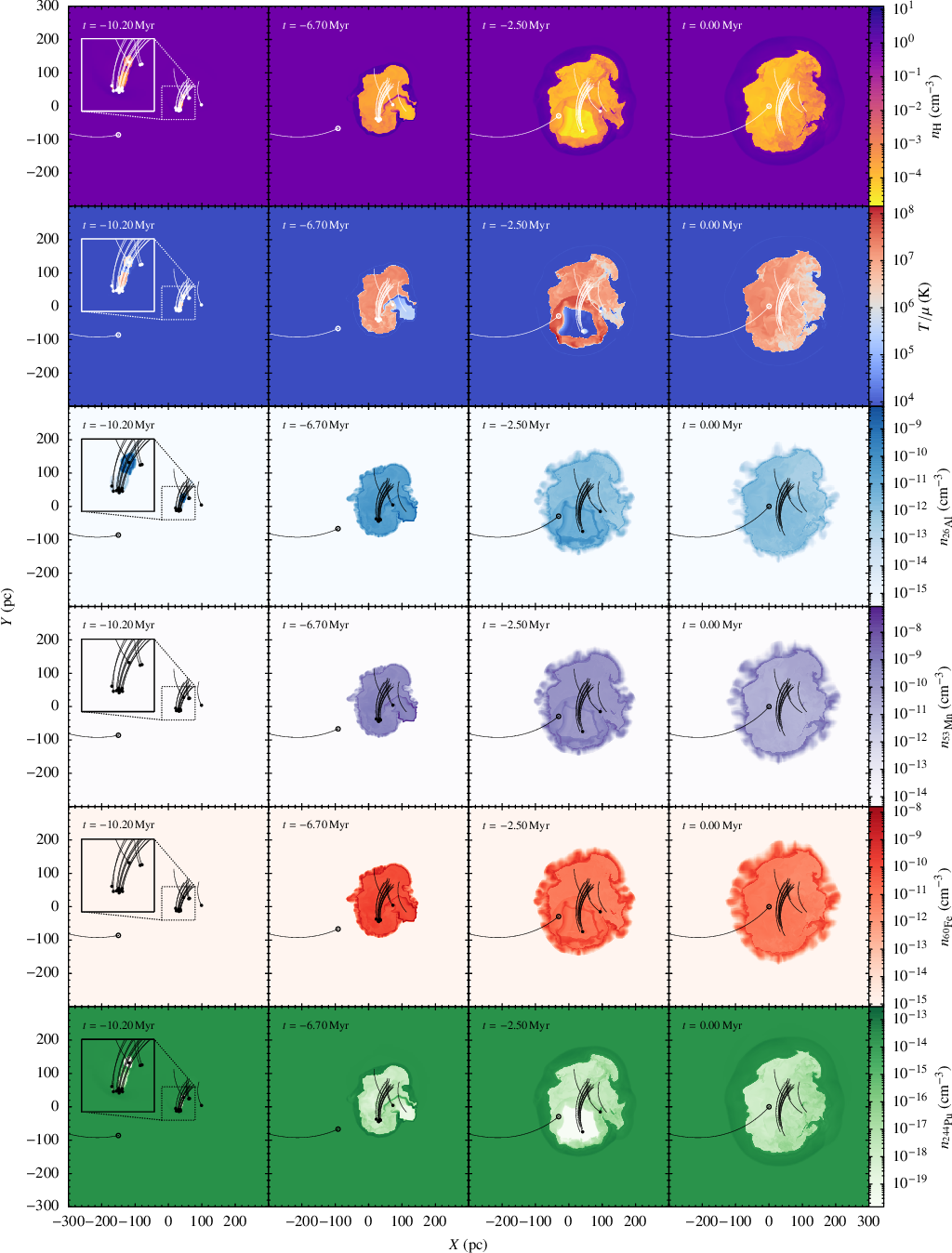}}
      \caption{As for Fig.~\ref{im:ts-sy}, but for slices at $Z=20.8$\,pc.
      }
         \label{im:ts-sz}
   \end{figure*}

      \begin{figure}
      \centering
   \resizebox{0.97\hsize}{!}
            {\includegraphics{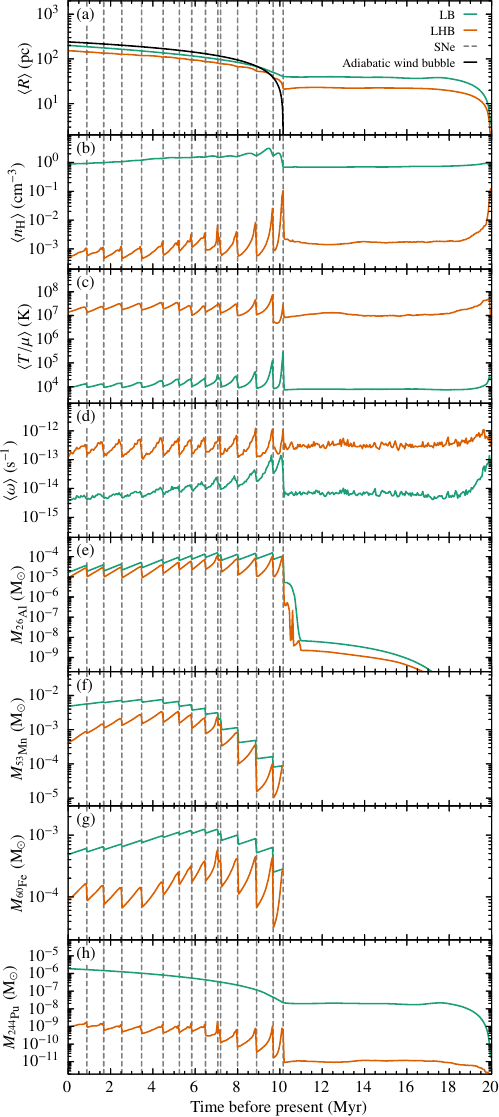}}
      \caption{Temporal evolution of (a) the equivalence radius (see text for the definition), (b) the average atomic hydrogen number density, (c) the average temperature ($T/\mu$), (d) the average absolute vorticity, (e) the \element[][26]{Al} mass, (f) the \element[][53]{Mn} mass, (g) the \element[][60]{Fe} mass, and (h) the \element[][244]{Pu} mass of the LB (turquoise curve) and the LHB (orange curve) gas in our fiducial simulation. The averages given for the LB (LHB) gas are weighted by mass (volume). The vertical dashed lines mark the explosion times of the individual SNe. Thus, the part of the profiles to the right of the rightmost dashed line reflects the conditions during the wind-driven phase, whereas almost all the rest of the profiles corresponds to the SN-driven phase. Only the part covering the last few 0.1\,Myr before present belongs to the final phase of evolution. In panel a, the self-similar wind solution of \citet{Weaver:1977} appropriate for the SN-driven phase is shown for comparison (black curve).}
         \label{im:evoprof}
   \end{figure}
   
Figures \ref{im:ts-sy}, \ref{im:ts-sz}, and \ref{im:ts_sx} show snapshots of the evolution of the LB taken at four different times. To generate the time profiles shown in Fig.~\ref{im:evoprof}, we had to locate the boundaries of the various bubbles in our simulations, for which we applied two strategies, similar to those exploited by \cite{Vasiliev:2017}. The first is based on the fact that any gas motion in the computational domain is due to flows associated with the formation of the LB (as the background medium is set up to be in equilibrium). We hence identified gas parcels as belonging to the LB if their flow speed exceeds a certain threshold. The value zero is not suitable for this in practice, as numerical errors always introduce small disturbances. We therefore chose $\unit[1]{km\,s^{-1}}$ instead. The turquoise lines in Fig.~\ref{im:evoprof} refer to this selection. The second strategy aims to find the gas shock-heated by stellar winds and SNe that fills the interiors of the bubbles. Since the temperature ($T$) of the ambient medium is kept within a narrow range around $\unit[10^4]{K}$, we assigned all cells containing gas with $T/\mu>\unit[10^5]{K}$ to the LHB, as represented by the orange lines in Fig.~\ref{im:evoprof}. The LHB is a true subset of the LB, compared to which it mainly lacks the cool dense shells. In order to specifically probe the conditions in the latter, the mean values calculated for the LB in panels b--d of Fig.~\ref{im:evoprof} are mass-weighted, whereas those for the LHB are volume-weighted. Although interstellar bubbles can never be perfectly spherical due to various reasons (see below), we estimated their sizes by introducing a (time-dependent) radius of equivalence, $\langle R\rangle=\sqrt[3]{3\,V/(4\,\pi)}$, taking for $V$ either the instantaneous volume of the LB (with radius $\langle R\rangle_\mathrm{LB}$) or that of the LHB (with radius $\langle R\rangle_\mathrm{LHB}$).

The LB in its entirety evolves along the three stages already identified by \cite{McCray:1987} and \cite{MacLow:1988} in their seminal papers for generic SBs in disc galaxies, namely a wind-driven phase, a SN-driven phase, and a final phase, which, depending on the initial and boundary conditions, is either a collapse or a `blowout' phase. From a purely gas-dynamical point of view, the first two phases are predominantly pressure-driven (i.e.~adiabatic), as the radiative cooling time of the hot bubble interiors is still well above their dynamical evolution timescale, so that the energy there is conserved. The final (or `snowplough') phase, on the other hand, is momentum-driven and initiated by the removal of the SB's interior pressure. The foundation for this is laid either by early fading SN activity, or a very dense or metal-rich environment, through which radiative cooling becomes important and the supershell collapses inwards under the influence of the external pressure and gravitational field (corresponding to the collapse phase). Or alternatively, when the polar caps of the supershell manage to accelerate away from the galactic mid-plane, either breaking through large vertical distances in the disc or breaking out of it altogether (corresponding to the blowout phase). Both situations are susceptible to the Rayleigh-Taylor (RT) instability that leads to the break-up of the supershell and thus allows for the pressure drop in the first place; for a semi-analytical solution in which the time dependence of this process is taken into account for the first time, see \cite{Schulreich:2022}.

\subsubsection{Wind-driven phase}
The birth of the LB takes place in principle already shortly after that of the massive stars in the Sco-Cen populations UCL/LCC and V1062~Sco at $t=\unit[-20]{Myr}$, which is also the start time of the simulations. These stars begin their lives in the main-sequence (MS) stage as early B-type stars. During this longest-lasting ($\sim$9--$\unit[17]{Myr}$ for the 14 LB progenitor stars in our model), hydrogen-burning phase of their evolution, the stars lose mass in the form of winds that are thought to be driven by momentum transferred from the stars' radiation field to metallic ions in their extended atmospheres. The time-averaged mass-loss rates are then on the order of $10^{-8}$ to $\unit[10^{-7}]{M_\sun\,yr^{-1}}$, and the wind speeds start at around $\unit[2900]{km\,s^{-1}}$ and gradually decrease with time to less than $\unit[2000]{km\,s^{-1}}$, with the mechanical luminosities ($\dot{M}\,\varv_\infty^2/2$) remaining approximately constant. 

The collision of such a fast wind with the surrounding ISM (whose isothermal sound speed, $c_{T,0}=\sqrt{\mathcal{R}\,T_\mathrm{ini}}$, is only about $\unit[7]{km\,s^{-1}}$) inevitably leads to the formation of a shock wave propagating away from the star. The pressure of the post-shock material causes the free-flowing fast wind to decelerate, introducing a second shock that propagates with the wind, albeit it is quasi stationary, often referred to as `termination shock'. The result is a complex double-shocked expanding structure separated by a contact discontinuity, as systematically studied by \cite{Weaver:1977}. Most of the volume of such a wind-blown bubble is occupied by the wind gas that went through the termination shock, establishing a hot, rarefied region. This cavity is enclosed by the shocked ISM, which is cooled and compressed into a thin, dense shell. 

The spherically symmetric picture just described assumes, in addition to a constant background medium, that the wind-blowing star is at rest relative to the ISM. Yet the progenitor stars of the LB have weakly supersonic space velocities in the range $\approx$9--$\unit[13]{km\,s^{-1}}$ along their tracks, as follows from our back-calculations (see Sect.~\ref{sec:MC_approach}). In such a case, the idealised flow pattern gets modified rather quickly, as the outer shock approaches a bow shock configuration determined by the balance between the ram pressure of the stellar wind and that of the ISM \citep[see e.g.][]{Weaver:1977,Wilkin:1996,Comeron:1998}. While this equilibrium applies to the leading side of the bubble, its trailing part continues to expand until the ram pressure it experiences falls below the thermal pressure of the ISM\footnote{The terms `leading' and `trailing' refer to the orientation with respect to the direction of motion of the star.}. As a result, this part of the shell collapses inwards and is fragmented by RT instabilities, allowing the hot interior gas to leak out and merge with the surrounding medium. What remains is just the leading part of the shell, which enters a stationary state, as the ISM density along the stellar trajectory, the space velocity of the star, and the wind's mechanical luminosity remain approximately constant. This allows for steady re-production of hot gas as fresh wind-released material continuously crosses the termination shock, passes around the free wind region, and gets trapped behind the star by the low-density channel carved into the ISM through the wind action. These channels are thus rough tracers for the trajectories of the wind-blowing stars.

Due to the small distance between the massive stars in our sample (about $\unit[38\pm 22]{pc}$ initially, and further decreasing with time; error indicates 1 standard deviation), most individual wind-blown bubbles already interact with each other during the first few hundred kyr, causing both their shells to merge and their MS-winds to collide. As evident from Fig.~\ref{im:evoprof}, this initially (note that time increases from right to left) heats the interior of the bubbles to extremely high temperatures (panel c) and causes relatively high compression (panel b). Also the mean absolute vorticity, $\langle\omega\rangle \equiv\langle|\vec{\nabla}\times\vec{u}|\rangle$ (panel d), experiences an early boost due to the high amount of turbulence introduced during the process, both in the bubble interiors and in the shells. The simultaneous rapid increase of the total volume covered by the expanding bubbles manifests itself in a superlinear growth of the effective radii (panel a).

At $t\approx\unit[-18.5]{Myr}$, the ram pressure experienced on average by the shells, $\rho_0\,\langle\dot{R}\rangle^2_\mathrm{LB}$ (dots refer to time derivatives), drops below the local ambient thermal pressure, $P_0\approx\unit[6168\,k_\mathrm{B}]{K\,cm^{-3}}$ (since all stars are located at low absolute Galactic heights, mid-plane values are fairly representative here). This is equivalent to the mean velocity of the shells undershooting the isothermal sound speed of the background medium, which allows the RT instability to fragment the trailing portions of the shells, as mentioned above. Thereupon, the average internal and external pressures almost equalise (local overpressures remain in the vicinity of both the termination shocks and the bow shocks), and the total volume that is instantaneously perturbed by the wind-blowing stars heads towards a steady state, which is reached at $t\approx\unit[-16]{Myr}$ at the latest. This is when $\langle R\rangle_\mathrm{LB}$ and $\langle R\rangle_\mathrm{LHB}$ (being measures for alternative parts of this volume) settle at plateau values of about $\unit[39]{pc}$ and $\unit[22]{pc}$, respectively (see Fig.~\ref{im:evoprof}a). It is important to note that since the structures are now no longer approximately spherical but rather cometary, these numbers are better understood as radii of curvature measured at the apsis of a hypothetical `average' stellar wind bow shock, with the axis of symmetry given by the mean space velocity vector of the massive stars.

With the initial expansion of the bubbles and the subsequent large-scale depressurisation, the temperature in the wind-shaped regions first decreases to then also become constant in time (see Fig.~\ref{im:evoprof}c). This is achieved on the one hand in the vicinity of the termination shock, since the temperature of the continuously shock-heated wind gas is so high that radiative cooling cannot change it, causing $\langle T/\mu\rangle_\mathrm{LHB}$ to stay at about $\unit[9.7\times 10^6]{K}$. On the other hand, when looking at the temperature averaged over the entire wind-shaped volume, $\langle T/\mu\rangle_\mathrm{LB}$, one finds that it is only slightly higher than that of the background medium, which, in view of the fact that the hot component just mentioned is contained in $\langle T/\mu\rangle_\mathrm{LB}$, means that there must be another component that is capable of depressing its value accordingly. This is the interstellar material that flows continuously along the shells after it has crossed the bow shock. It can cool efficiently (though not to below $T_\mathrm{ini}$ in our simulations, as to keep the stratified background medium stable), which is the reason for the shells being rather thin and dense. The values on which $n_\mathrm{H}$ and $\omega$ settle on average are $\unit[1.7\times 10^{-3}]{cm^{-3}}$ and $\unit[3.1\times 10^{-13}]{s^{-1}}$ for the hot gas component, and $\unit[0.7]{cm^{-3}}$ and $\unit[6.8\times 10^{-15}]{s^{-1}}$ for the entire wind-shaped volume (Fig.~\ref{im:evoprof}b,d), though, as with all gas properties, with rather high spatial deviations.

Another consequence of the shells assuming a bow shock configuration is that the mass of the pre-seeded \element[][244]{Pu} that is instantaneously entrained approaches a constant value ($\approx$$\unit[2\times 10^{-8}]{M_\sun}$; turquoise curve in Fig.~\ref{im:evoprof}h), already including the (very inefficient) radioactive decay. The bow shock namely acts like the shovel of a snowplough, which, assuming a uniformly snow-covered road and constant driving speed, pushes a pile of snow in front of it, which though consisting always of a different combination of snowflakes (i.e.~\element[][244]{Pu} atoms), has a mass that remains constant over time (after the plough has already travelled far enough for this stationarity to be achieved). This is in contrast to an expanding spherical shock, where the entrained mass must always keep increasing because it has no possibility of escaping in any direction. For the LB, the latter situation corresponds to the early wind-driven phase, in which the shells of the bubbles are still fully intact (rightmost part of Fig.~\ref{im:evoprof}). A flatter but similarly shaped profile is obtained for the \element[][244]{Pu} mass belonging to the hot gas component at a given time (orange curve in Fig.~\ref{im:evoprof}h): While it initially increases because the shells are still expanding and closed so that they can keep their interior at high temperatures, the curve flattens out as soon as most of the hot gas escapes and only that \element[][244]{Pu} enriched LISM can remain hot which has crossed the bow shock and mixed with the wind-shocked material. However, this mass amounts to only about $\unit[10^{-11}]{M_\sun}$.

Of all the radioisotopes considered in this work, only \element[][26]{Al} is released during the wind-driven phase. Hence, with the ongoing wind action, its mass in the LISM grows steadily, especially within the hot wind-shocked regions (see Fig.~\ref{im:evoprof}e). The pronounced boost both curves experience at $t\approx\unit[-11]{Myr}$ marks the time at which the most massive star of our model is the first to enter the red supergiant (RSG) phase, into which all others will follow in order of descending initial mass. In this phase, which lasts a few hundred kyr to a few Myr, the stars increase immensely in size and begin to burn helium in their cores. The large sizes imply much smaller escape velocities off their surfaces and thus smaller wind speeds. The mass-loss rates, however, are quite high, ranging on average from $\sim$$10^{-6}$ to a few $\unit[10^{-5}]{M_\sun\,yr^{-1}}$. RSG winds are thought to be driven by radiation pressure on dust grains.

For stars with initial masses less than $\unit[20]{M_\sun}$, evolution ends in the RSG stage. Higher-mass stars (two in our model; see Table \ref{tab:input}) shed their hydrogen envelopes and evolve further into WR stars, after they have moved back to the blue part of the HRD. This comparatively low WR mass limit \citep[][e.g., obtained 32\,M$_\sun$ via their non-rotating models]{Schaller:1992}, and also the blueward evolution, is due to the enhanced mass loss during the RSG phase present in the rotating models of \cite{Ekstroem:2012}, which the authors attribute to the star's luminosity becoming supra-Eddington in its outer layers. WR stars lose mass in the form of radiation-driven winds, with mass-loss rates at the lower end of their RSG predecessors, but exhibiting wind speeds that are more than ten times higher ($\sim$$\unit[10^3]{km\,s^{-1}}$). After a few hundred kyr, however, this last phase of evolution is also over.

Purely in terms of the wind-driven phase, we can therefore summarise that it is the very slow but dense RSG-winds that have the highest contribution to the total amount of \element[][26]{Al} released, whereas the subsequent high-momentum WR-winds as well as the MS-winds blown simultaneously by neighbouring massive stars are mainly responsible for its large-scale (turbulent) dispersal. It is also the latter winds that, essentially, set the size of the wind-shaped volume \citep[cf.][]{Dwarkadas:2007}.

For the range of initial masses covered by CC SN progenitors, the total mass lost to winds increases with the initial mass of the star. For our LB progenitors, this is between about 19 and 64 per cent of their respective initial masses. Also the cumulative wind-released \element[][26]{Al} mass increases with the initial mass of massive stars and amounts to between a few $\unit[10^{-8}]{M_\sun}$ and a few $\unit[10^{-6}]{M_\sun}$ for our sample. In total, this is about $\unit[3.6\times 10^{-5}]{M_\sun}$ of \element[][26]{Al} released by all 14 LB progenitors over their entire lifetime, which is close to the amount ejected when the two lowest-mass stars in our sample go SN (see Table~\ref{tab:input}). The energy liberated by all LB progenitors through their life via winds totals about $\unit[4.1\times 10^{50}]{erg}$, which is only $\sim$41 per cent ($\sim$3 per cent) of the energy released by a single (all) SN(e) in our model.

The maps in the first column of Figs.~\ref{im:ts-sy}, \ref{im:ts-sz}, and \ref{im:ts_sx} show cuts through the wind-shaped regions as they appear immediately before the first SN explodes (i.e.~at the end of the wind-driven phase), together with the projected positions of the massive stars (filled circles) and the Sun (empty circle). Since the bulk of UCL/LCC and V1062~Sco, including the LB progenitor stars, has moved almost parallel to the $Y$-axis (coming from higher $Y$-values) since formation and simultaneously crossed the Galactic mid-plane from south to north, as can be seen from the trajectories (thin solid lines), the majority (10 out of 14) of the individually blown channels could coalesce into a prominent obliquely elongated basin of hot gas extending from the instantaneous position of the massive stars, which is close to the coordinate origin, mainly in the positive $Y$ and negative $Z$ directions. In close proximity are the similarly aligned but much narrower wind-shaped regions of the other LB progenitors, two driven by single stars and one driven by two very closely spaced stars. None of them has yet reached the $YZ$ cutting plane.

The fact that the $Y$-axis is always tangential to the direction of Galactic rotation in which our computational domain is thought to participate (so that we can assume the LISM to be at rest) implies that the formation of the wind-shaped regions should not have been significantly affected by the shear motion due to the differential rotation of the Galaxy, as has already been pointed out by \cite{Fuchs:2006}. And since the stars maintain their direction of motion to a good approximation up to the present time, we decided not to include Galactic shear in our simulations at all. We actually believe that the absence of Galactic shear has been a crucial factor in allowing the stellar feedback processes that took place over the long period of $\unit[20]{Myr}$ to combine to form the LB in the first place, rather than drifting too far apart beforehand.

\subsubsection{Supernova-driven phase}
The most massive star in our model, which is also the first to explode as a SN, at $t\approx\unit[-10.16]{Myr}$, belongs to one of the two single star-driven, and thus smallest, bow shock regions. The blast wave generated by the explosion can expand freely due to the low density in the wind-blown channel. First it hits the shell behind the bow shock, as this is closest to the explosion centre, whereupon a reflected shock runs back into the remnant. The shock transmitted into the shell remains temporarily trapped within it and becomes radiative. Simultaneously, both the section of the blast wave that runs into the trailing side of the channel and the reflected shock that also propagates there, but lags behind the blast wave, undergo an oblique interaction with the lateral channel walls, compressing them. Upon impact at the rear end of the channel, the blast wave also splits into a transmitted and a reflected component. The transmitted shock can cross the channel boundary almost unhindered before propagating adiabatically into the pristine LISM. The reflected shock, on the other hand, runs back into the cavity and shortly afterwards merges with the shock that was reflected at the bow shock shell, which greatly enhances the turbulence of the flow within the cavity. Ultimately, both the radiative shocks transmitted into the bow shock shell and into the lateral channel walls succeed in breaking out into the LISM. 

Their undisturbed propagation there, however, soon comes to an end, namely shortly after $t\approx\unit[-9.68]{Myr}$, the time at which the next SN, now in the bow shock region densely populated by massive stars, explodes with the result that the remnant there undergoes an evolution analogous to that of the first. The two remnants collide and subsequently merge completely. The resulting remnant is not spherical but elongated more parallel than perpendicular to the Galactic disc due to the mainly horizontally displaced explosion centres and the preferential energy deposition in the leading and trailing parts of the former wind-shaped regions, as the SN shocks strike there at lower inclination angles and are thus particularly strong. The dense shell with which the remnant surrounds its tenuous interior is heavily corrugated, at this stage mainly due to thermal instabilities.

All further SNe, with the exception of the sixth, take place within this expanding cavity. The progenitor star of the sixth SN is the only one that does not belong to UCL/LCC but to V1062~Sco. Because of this, it has always been somewhat isolated from the other massive stars. In fact, it constitutes the last remaining stellar wind bow shock until its death, at $t\approx\unit[-7.07]{Myr}$, with the massive stars in the interior of the LB not being able to do so because of the high temperatures and thus sound speeds prevailing there. The explosion of the V1062~Sco star still happens so close to the cumulative shell of all previous SNe that it is hit very hard by the radiative shock of the young isolated remnant. The effective impulsive acceleration experienced by the contact surface that is established between the two colliding shells gives rise (together with the density gradient across it) to the Richtmyer-Meshkov instability, which fragments the layer of interaction almost instantly, opening up a hole. Through this hole, the hot gas content of the young remnant, as well as the layer fragments and turbulent eddies arising from the collision, are sucked into the interior of the LB, while the remaining shell of the shattered remnant gets incorporated into the further expanding supershell. This moment is captured in the snapshots in the second column of Figs.~\ref{im:ts-sy}, \ref{im:ts-sz}, and \ref{im:ts_sx}.

Figure \ref{im:evoprof} shows the detailed effect of each SN on both the LB as a whole and exclusively on its hot interior, with the explosion times of the SNe marked by vertical dashed lines. As can be observed in panel a, the enormous energy release of the first SN already causes the growth of the effective radii to be reignited after they have settled to plateau values in the wind-driven phase, with the expansion of the LB gas proceeding continuously, whereas that of the LHB gas, which is more immediately affected by the explosions, occurs in distinct bursts. Since the volumes on which these radii are based are now almost always contiguous, in contrast to the wind-driven phase where this was never the case, the difference in the radii provides a rough estimate of the LB's actual shell thickness.

That the latter can also grow with time despite efficient radiative cooling is due to the fact that the shell decelerates except for the short, pulse-like periods of acceleration in the aftermath of SNe. In the rest frame of the already corrugated contact discontinuity, which separates the ejected material heavily compressed by the regular SN activity from the thinner swept-up LISM, the acceleration vector is thus for most of the time opposite and over much of the interface skewed to the density gradient vector, which is a configuration prone to the RT instability. The outward growing RT fingers deform the outer shock, leading to the increase in layer thickness. This is best observed in the maps of the number densities of \element[][26]{Al}, \element[][53]{Mn}, and \element[][60]{Fe} (panels in the third to fifth rows of Figs.~\ref{im:ts-sy}, \ref{im:ts-sz}, and \ref{im:ts_sx}), as the background medium has been set up to be devoid of these radioisotopes, making the shell volume composed of the swept-up LISM effectively translucent.

As demonstrated by \cite{MacLow:1988}, SBs in the SN-driven phase can be described approximately as very large wind-blown bubbles and therefore by means of the self-similar expansion law found by \cite{Weaver:1977}, where the stellar wind emanating from a single star is replaced by the SNe occurring shortly after each other, close together, as a continuous source of internal energy. That this simple model also works quite well for the LB can be shown by taking the ratio between the explosion energy of a single SN and the average time span between successive SNe, $\Delta t_\mathrm{exp}\approx \unit[0.71]{Myr}$, as a mechanical luminosity equivalent, that is $L_\mathrm{SN}=E_\mathrm{SN}/\Delta t_\mathrm{exp}\approx \unit[4.44\times 10^{37}]{erg\,s^{-1}}$. With this, the instantaneous radius of the LB can be calculated via
\begin{equation}
\begin{split}
R &=\left(\frac{125}{154\,\pi}\right)^{1/5}\,\left(\frac{L_\mathrm{SN}}{\rho_0}\right)^{1/5}\,t_\mathrm{SN}^{3/5}\\
&\approx \unit[241]{pc}\,\left(\frac{L_\mathrm{SN}}{\unit[4.44\times 10^{37}]{erg\,s^{-1}}}\right)^{1/5}\,\left(\frac{n_{\mathrm{H},0}}{\unit[0.7]{cm^{-3}}}\right)^{-1/5}\\
&\quad\times \left(\frac{t_\mathrm{SN}}{\unit[10.16]{Myr}}\right)^{3/5}\,,
\end{split}
\label{eq:weaver}
\end{equation}
where $t_\mathrm{SN}$ denotes the time span since the first SN, and for the ambient density neither that within the wind-shaped region nor the vertical Galactic gradient is used but simply the density value in the mid-plane, since this is, to a good approximation, what most parts of the supershell `feel' over the longest time of their expansion. Equation (\ref{eq:weaver}) is plotted as a black curve in Fig.~\ref{im:evoprof}a. As can be seen, the radius advantage of the numerical solution, originating from the wind-driven phase, is made up very quickly, with $\langle R\rangle_\mathrm{LB}$ aligning with the $t_\mathrm{SN}^{3/5}$-scaling from about the eighth SN on. That the analytical solution then still slightly overestimates $\langle R\rangle_\mathrm{LB}$, by a factor of 1.25 or so, is not surprising and is mainly due to the fact that (1) both the ambient pressure and gravitational acceleration are non-zero, (2) the supershell suffers from radiative cooling losses, (3) the supershell consists not only of the swept-up LISM but also of the material ejected by the SNe and the stellar winds (further increasing its mass and thus inertia), and (4) discrete SN events at the given rate correspond only very roughly to a continuous wind.

Apart from driving the expansion of the LB, each individual SN blast wave causes its interior to be abruptly compressed and heated up on average. In the long run, however, the average density of the LHB (and also of the LB) gas decreases due to the expansion. The shock heating, on the other hand, is able to counteract the adiabatic cooling, so that the LHB gas remains similarly hot on average over long periods of time\footnote{A similar argument was put forward by \cite{Kahn:1998}, who showed analytically that if successive explosions are delayed sufficiently long for radiative cooling to set in, the expansion of a SB can be substantially enhanced.} (see Fig.~\ref{im:evoprof}b,c) -- a temperature minimum is effectively set by the artificial cooling floor. The outward propagation of the SN blast waves, which are highly deformed due to the density inhomogeneities they encounter, and their subsequent asymmetric reflection from the likewise corrugated contact discontinuity in the supershell causes the mean vorticity in the LB to shoot up as well. The level of vorticity can also be maintained in the long term, especially in the LHB gas, as this is almost permanently struck by reflected shocks (see Fig.~\ref{im:evoprof}d). 

The SNe also enrich the LB episodically with radioisotopes, in our model with \element[][26]{Al}, \element[][53]{Mn}, and \element[][60]{Fe}, as clearly visible from the mass profiles in Figs.~\ref{im:evoprof}e--g. The variation of the absolute peak heights reflects the initial mass- and thus explosion time-dependent yields of the radioisotopes (see columns 10--12 of Table \ref{tab:input}). The reason why the post-peak declines for a given isotope are not the same for the LB and LHB gas, is that the former also contains the supershell and it is there where the bulk of the radioisotopes are dumped (see the panels in the third to fifth row of Figs.~\ref{im:ts-sy}, \ref{im:ts-sz}, and \ref{im:ts_sx}). Accordingly, the mass drops of \element[][26]{Al}, \element[][53]{Mn}, and \element[][60]{Fe} in the highly diluted LHB gas are due to radioactive decay and loss of material to the supershell, which is why they are always steeper than in the LB gas, which is exclusively affected by decay. In contrast, the mass of \element[][244]{Pu}, the only radioisotope in our model that is not released by SNe, increases steadily because its half-life is very long compared to the timescale over which it is swept up by the supershell, with not insignificant amounts of \element[][244]{Pu} being transferred from the LHB to the LB gas by propagating SN shocks (see Fig.~\ref{im:evoprof}h).

   \begin{figure}
   \resizebox{\hsize}{!}
            {\includegraphics{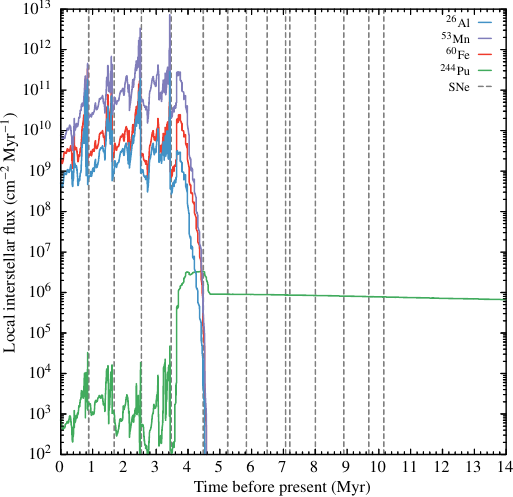}}
      \caption{Local interstellar fluxes of the four radioisotopes specified in the legend as a function of time before present, as derived from our fiducial simulation. The vertical dashed lines mark the explosion times of the individual SNe.}
         \label{im:fluxplot}
   \end{figure}

At $t\approx\unit[-4.64]{Myr}$, the Sun crosses the outer shock of the LB, thus entering the swept-up LISM. This can be seen most accurately from the sudden rise in the local interstellar flux of \element[][244]{Pu}, as represented by the green curve in Fig.~\ref{im:fluxplot}. About $\unit[60]{kyr}$ later, the SS crosses the contact discontinuity in the supershell, which is noticeable from the subsequent increase in the fluxes of the SN-released radioisotopes \element[][26]{Al} (blue curve), \element[][53]{Mn} (purple curve), and \element[][60]{Fe} (red curve) -- less abruptly than for \element[][244]{Pu}, however, since the RT instability provides for mixing within the supershell. The Sun's departure from the supershell or, equivalently, its arrival in the LHB volume, at $t\approx\unit[-3.65]{Myr}$, is indicated by a steep decline in each of the four radioisotopic fluxes. All pulses that appear in the diagram afterwards are generated either by the SN blast waves themselves (these peaks then lie extremely close to the dashed lines that mark the SN explosion times as in Fig.~\ref{im:evoprof}) or by their reflected shocks. The signals generated by the latter are usually much weaker because there is hardly any material left to sweep up after the blast wave has passed through. For \element[][26]{Al}, \element[][53]{Mn}, and \element[][60]{Fe}, the shape of the profile over the residence time of the Sun in the LHB volume is determined primarily by the SN yields, and to a much lesser extent by the radioactive decay. For \element[][244]{Pu}, on the other hand, it is exclusively the radioactive decay. This however proceeds so slowly for \element[][244]{Pu} that its local interstellar flux seems to be roughly constant on a time-average basis. 

The maps in the third column of Figs.~\ref{im:ts-sy}, \ref{im:ts-sz}, and \ref{im:ts_sx} capture the blast wave of that 12th SN about $\unit[40]{kyr}$ after explosion, which is mainly responsible for the measured anomalies of \element[][60]{Fe} and \element[][53]{Mn} (see Sect.~\ref{sec:compmeas}). The pockets of LHB gas, as they appear in the XZ slice plane (Fig.~\ref{im:ts-sy}), result from the supershell's anisotropic expansion due to inhomogeneities in the background medium that were already introduced during the wind-driven phase.

\subsubsection{Approaching the final phase}
At $t\sim\unit[-0.5]{Myr}$, the northern polar cap of the LB enters a phase of acceleration that lasts until today and will probably continue beyond. Less than a Myr before this time, the upper tip of the supershell has also passed twice the scale height of the stratified background medium model, being $H\approx\unit[119]{pc}$. Both aspects are associated with the LB beginning to expand preferentially perpendicular to the Galactic mid-plane \citep[cf.][]{MacLow:1988,Roy:2013}, leading to its vertical elongation and making the spherically symmetric bubble approximation of \cite{Weaver:1977} an increasingly poor fit -- all the more so since by $t\approx\unit[-0.88]{Myr}$ the energy input from the SNe (and stellar winds) runs out. The reason why the northern part of the LB experiences this acceleration first is that all SNe have occurred at positive Galactic heights (see fourth column of Table \ref{tab:input}), so their blast waves experience the least resistance from the background medium in that direction.

However, at the end time of the simulation, which corresponds to the present time, the RT instability in the shell has not yet progressed to the point where it could break it up to allow the hot interior gas to leak out and the interior pressure to drop significantly (see maps in the fourth column of Figs.~\ref{im:ts-sy}, \ref{im:ts-sz}, and \ref{im:ts_sx}, and also Fig.~\ref{im:evoprof}). Accordingly, the LB has not yet entered the final phase of its evolution, which will most likely be of blowout-type, but is on the verge of doing so. 

To provide an even better impression of the structure of the LB at $t=0$, we show a volume rendering of the atomic hydrogen number density in Fig.~\ref{im:nhrendering}. Table \ref{tab:final} gives a summary of the gas-dynamical properties of both the LB and the LHB at this time. The thermal pressure we find on average for the present LHB gas, $\langle P/k_\mathrm{B}\rangle_\mathrm{LHB}=\langle n_\mathrm{H}/X\rangle_\mathrm{LHB}\,\langle T/\mu\rangle_\mathrm{LHB}\approx \unit[10\,100]{K\,cm^{-3}}$ (with $X\approx 0.707$ being the hydrogen mass fraction), agrees well with the value of $\unit[10\,700]{K\,cm^{-3}}$ estimated by \cite{Snowden:2014} through a combination of disparate observational results.

\begin{table}
\caption{Present-day properties of the LB and LHB gas, as derived from our fiducial simulation. Errors indicate 1 standard deviation.} 
\label{tab:final}      
\centering                        
\begin{tabular}{l c c}           
\hline\hline  
 & LB & LHB\\
\hline 
$\langle R\rangle$ (pc) & $201$ & $153$\\
$n_\mathrm{H}$ (cm$^{-3}$) & $0.87\pm 0.30$ & $(5.1\pm 16.6)\times 10^{-4}$\\
$T/\mu$ (K) & $(9.0\pm 194.0)\times 10^3$ & $(1.4\pm 0.9)\times 10^7$\\
$\omega$ (s$^{-1}$) & $(3.8\pm 9.8)\times 10^{-15}$ & $(1.3\pm 1.3)\pm 10^{-13}$\\
$M_{\element[][26]{Al}}$ (M$_\odot$) & $1.6\times 10^{-5}$ & $9.0\times 10^{-6}$\\
$M_{\element[][53]{Mn}}$ (M$_\odot$) & $4.8\times 10^{-3}$ & $4.2\times 10^{-4}$\\
$M_{\element[][60]{Fe}}$ (M$_\odot$) & $4.8\times 10^{-4}$ & $8.9\times 10^{-5}$\\
$M_{\element[][244]{Pu}}$ (M$_\odot$) & $1.9\times 10^{-6}$ & $9.5\times 10^{-10}$\\
\hline                                   
\end{tabular}
 \tablefoot{
   $\langle R\rangle$: equivalence radius (see text for the definition); $n_\mathrm{H}$: atomic hydrogen number density; $T/\mu$: temperature (per mean molecular weight); $\omega$: absolute vorticity; $M_{\element[][26]{Al}}$: \element[][26]{Al} mass; $M_{\element[][53]{Mn}}$: \element[][53]{Mn} mass; $M_{\element[][60]{Fe}}$: \element[][60]{Fe} mass; $M_{\element[][244]{Pu}}$: \element[][244]{Pu} mass. 
   }

\end{table}

\subsection{Comparison with radioisotopic measurements}
\label{sec:compmeas}

   \begin{figure*}
   \resizebox{\hsize}{!}
            {\includegraphics{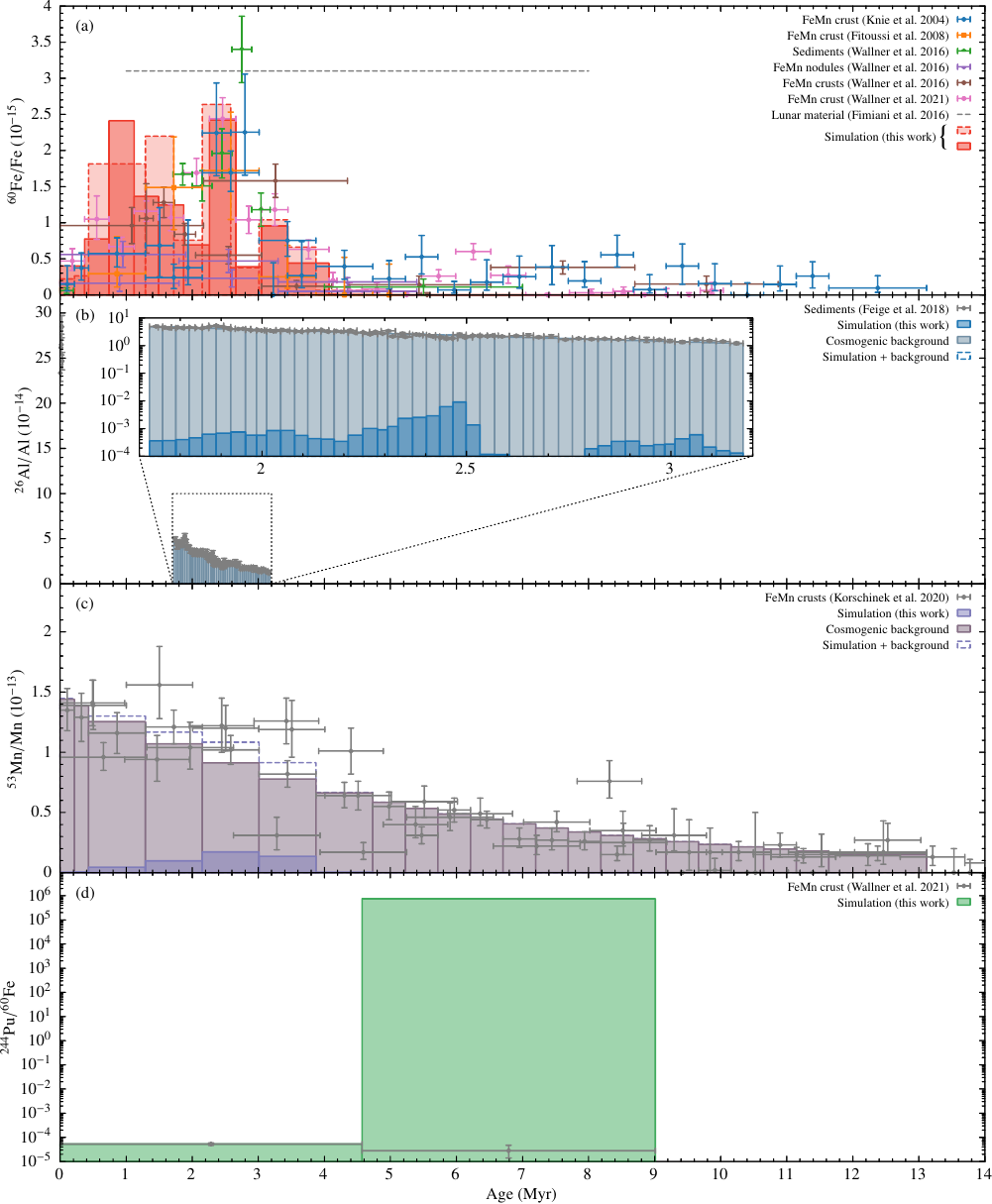}}
      \caption{Comparison of the measured (symbols with error bars and dashed grey line) and modelled (histograms) ratios (a) $\element[][60]{Fe}/\element[][]{Fe}$, (b) $\element[][26]{Al}/\element[][]{Al}$, (c) $\element[][53]{Mn}/\element[][]{Mn}$, and (d) $\element[][244]{Pu}/\element[][60]{Fe}$ from our fiducial simulation as a function of age of the respective reservoir. The inlay in panel b shows a magnification of the boxed area. All data, except $\element[][244]{Pu}/\element[][60]{Fe}$, are not corrected for radioactive decay. The crust ages of \cite{Knie:2004} and \cite{Fitoussi:2008} have been updated to account for the latest half-life of \element[][10]{Be} \citep[$t_{1/2,\element[][10]{Be}}\approx 1.387$\,Myr;][]{Korschinek:2010} used for the dating. For the \cite{Fimiani:2016} lunar data included in panel a, which for reasons of micrometeoritic gardening on the lunar regolith cannot be resolved better than about 8\,Myr in time, we use the mean value determined by \cite{Ertel:2023}.}
         \label{im:crustplot}
   \end{figure*}

Figure \ref{im:crustplot} shows how the local interstellar fluxes obtained from our fiducial simulation and converted into density ratios according to the procedure described in Sect.~\ref{sec:radioiso} compare with the measured data, assuming for each radioisotope the same survival fraction of $f_i\equiv f=0.0004$. This value provides for our current calculations the best fit with the measurements and is consistent with the finding that around 1 per cent of the radioisotope-bearing dust at the heliosphere boundary can enter the SS \citep{Fry:2015}, and of this again only a fraction whose estimates range from a few per cent to a few tens of per cent \citep{Wallner:2016,Wallner:2021} can ultimately be incorporated into the geological archive.

To make further comparisons with measurements, we plotted the time profiles for the simulated local interstellar flux ratios $\element[][60]{Fe}/\element[][26]{Al}$ and $\element[][53]{Mn}/\element[][60]{Fe}$ in Fig.~\ref{im:ratiosplot}.

\subsubsection{$\element[][60]{Fe}/\element[][]{Fe}$}
Both histograms shown in Fig.~\ref{im:crustplot}a derive from the same simulation, but differ in which FeMn crust sample was used for the time binning (as the local interstellar flux is smeared over time intervals, each with different boundaries and sometimes widths resulting from the sample cutting). In the case of the dashed histogram, this is the crust 237KD studied by \cite{Knie:2004}, whereas the solid histogram is based on Crust-3 analysed by \cite{Wallner:2021}. Both crusts stem from the Pacific Ocean. The mean number densities of stable Fe ($m_{\element[][]{Fe}}\approx\unit[55.85]{g\,mol^{-1}}$) are very similar in both crusts, namely $n_{\element[][]{Fe}}\approx\unit[2.28\times 10^{21}]{cm^{-3}}$ (since $w_{\element[][]{Fe}}\approx 0.15$ and $\rho_\mathrm{r}\approx\unit[1.40]{g\,cm^{-3}}$) for 237KD and $n_{\element[][]{Fe}}\approx\unit[2.86\times 10^{21}]{cm^{-3}}$ (since $w_{\element[][]{Fe}}\approx 0.14$ and $\rho_\mathrm{r}\approx\unit[1.90]{g\,cm^{-3}}$) for Crust-3. In both cases, the influx becomes significant at $\sim$$\unit[4]{Myr}$ ago, corresponding roughly to the Sun's passage through the LB's contact discontinuity, which is consistent with the measured data (dots with error bars). Likewise, the subsequent gradual increase in the measured $\element[][60]{Fe}/\element[][]{Fe}$ ratio is satisfactorily reproduced by the simulation. The local maximum around $\unit[2.5]{Myr}$ ago is perfectly replicated by both histograms, both in terms of position and height. The further progression of the measured data is also by and large well reproduced, considering how many hard to constrain parameters the LB model is based on. The signal caused primarily by the last SN in our model $\unit[0.88]{Myr}$ ago, which manifests itself as a local maximum in the finer-binned Crust-3 of \cite{Wallner:2021}, is not observed in this form, though. This may indicate that (1) the time at which the last SN in the LB should have occurred, as derived from the NEI plasma simulations of \cite{Avillez:2012}, which sets the value of $M_\mathrm{ion}$ and thus affects the IMF mass binning as a whole, may be somewhat further back in the past, and/or that (2) the most probable mass distribution of the LB progenitor stars may not necessarily be that which exactly underlay the formation of the LB, although it probably did not deviate greatly from it. We note that if one simply omits the last SN (by taking its progenitor star out of the simulation), this flaw in the modelled profile is also removed (see Fig.~\ref{im:crustplotapp}).

As already indicated by the measurements contained in Fig.~\ref{im:crustplot}a, the \element[][60]{Fe} influx onto Earth continues up to the present time, albeit significantly reduced compared to its peak. This is corroborated by measurements of Antarctic snow \citep{Koll:2019} and deep-sea sediment samples \citep{Wallner:2020}, covering the age ranges 0--\unit[20]{yr} and 0--\unit[33]{kyr}, respectively. The ratio of the mean flux during the time range spanned by the younger \element[][60]{Fe} peak (1.7--\unit[3.2]{Myr} ago) to the recent flux is between about 7 (sediment) and 20 (snow), with averaging the recent \element[][60]{Fe} fluxes from table 2 of \citet{Wallner:2020} yielding a value of $\sim$10. We averaged the local interstellar \element[][60]{Fe} flux from our simulation for the three time ranges just mentioned and obtained $\unit[1.99\times 10^{9}]{cm^{-2}\,Myr^{-1}}$ for $t=0$ (since the time interval spanned by Antarctic snow is smaller than our simulation time steps, we simply picked the present-day value here), $\unit[1.79\times 10^{9}]{cm^{-2}\,Myr^{-1}}$ for 0--\unit[33]{kyr} ago, and $\unit[1.85\times 10^{10}]{cm^{-2}\,Myr^{-1}}$ for 1.7--\unit[3.2]{Myr} ago. The corresponding flux ratios between the younger peak and the recent time intervals are thus 9.3 (snow) and 10.4 (sediment), which agree well with the measured ratios.

\subsubsection{$\element[][26]{Al}/\element[][]{Al}$}
In order to meaningfully compare the modelled $\element[][26]{Al}/\element[][]{Al}$ ratio with the one measured in deep-sea sediment cores from the Indian Ocean \citep[ELT 45-21 and ELT 49-53;][]{Feige:2018}, the high cosmogenic background of \element[][26]{Al} must be taken into account. We estimated the latter by binning the decay curve of \element[][26]{Al} for the samples of ELT 53-49 while applying for the initial concentration the mean value of the uppermost layers, being $(\element[][26]{Al}/\element[][]{Al})_0\approx 2.56\times 10^{-13}$, where $n_\mathrm{\element[][]{Al}}\approx\unit[2.51\times 10^{19}]{cm^{-3}}$. By doing so, we implicitly assumed that (1) the top layer is not or only negligibly contaminated by stellar feedback processes, and (2) the influx of \element[][26]{Al} not related to stellar feedback was constant over the time range covered by the samples ($\approx$1.72--$\unit[3.18]{Myr}$ ago). The result is the grey-blue filled histogram in Fig.~\ref{im:crustplot}b, which, as better seen in the magnification, already perfectly fits the measurements. It can therefore be expected that the \element[][26]{Al} background is simply too high for contributions from stellar feedback processes to stand out from it, as already concluded by \cite{Feige:2018}. We can confirm this speculation for the first time in this paper quantitatively on the basis of our simulation results (deep blue filled histogram). If these are added to the background values, the outcome, the dashed empty histogram, does not noticeably exceed the background, even under magnification.

\subsubsection{$\element[][53]{Mn}/\element[][]{Mn}$}
For \element[][53]{Mn}, which also has a high cosmogenic background, we proceeded analogously to \element[][26]{Al}. As initial concentration we took the mean of the $\element[][53]{Mn}/\element[][]{Mn}$ values in the uppermost layer of the four FeMn crusts examined by \cite{Korschinek:2020}, being $(\element[][53]{Mn}/\element[][]{Mn})_0\approx 1.47\times 10^{-13}$. For the binning, we used 237KD as it covers the longest uninterrupted time range of all four ($\approx$0--$\unit[13.12]{Myr}$ ago). The crust's mean number density of stable Mn is $n_{\element[][]{Mn}}\approx\unit[3.50\times 10^{21}]{cm^{-3}}$ (since $w_{\element[][]{Mn}}\approx 0.23$, $m_{\element[][]{Mn}}\approx \unit[54.94]{g\,mol^{-1}}$, and $\rho_\mathrm{r}\approx\unit[1.40]{g\,cm^{-3}}$). The histogram that emulates the background is the grey-purple filled one in Fig.~\ref{im:crustplot}c. On top of this we put the simulation data (deep purple filled histogram) resulting in the dashed empty histogram. As can be seen, it is consistent with the measurements of \cite{Korschinek:2020} both in terms of the deposition time of the largest amount of \element[][53]{Mn} ($\sim$$\unit[2.5]{Myr}$ ago) and the absolute $\element[][53]{Mn}/\element[][]{Mn}$ ratio, which is $\sim$$1.6\times 10^{-14}$ above that expected from cosmogenic production at that time.

\subsubsection{$\element[][244]{Pu}/\element[][60]{Fe}$}
The two data points plotted in Fig.~\ref{im:crustplot}d show the decay-corrected radioisotopic ratio $\element[][244]{Pu}/\element[][60]{Fe}$ measured and averaged over a total of three layers of Crust-3, taken from \cite{Wallner:2021}. As described in Sect.~\ref{sec:radioiso}, we used the data point spanning the interval $0$--$\unit[4.57]{Myr}$ ago and thus the residence time of the SS in the LB to calibrate the $\element[][244]{Pu}$ concentration in our simulations. It is thus not surprising that the corresponding bin of the simulation-derived histogram agrees perfectly with the measurement value. What is more interesting is how well (or badly) the other measurement value and histogram bin match (both spanning the time range $4.57$--$\unit[9.01]{Myr}$ ago). Here it turns out that the simulation overestimates the measurement by an enormous ten orders of magnitude. As will be discussed further in Sects.~\ref{sec:neighbsbs} and \ref{sec:stimings}, we see this large discrepancy as a strong indication that the Sun, before entering the LB, must have passed through a medium that, similar to the volume of the LB, must have been almost empty of $\element[][244]{Pu}$ atoms and thus presumably of gas per se.

   \begin{figure}
   \resizebox{\hsize}{!}
            {\includegraphics{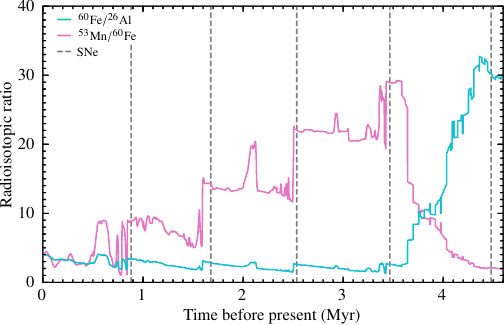}}
      \caption{Selection of ratios of the local interstellar fluxes of the radioisotopes \element[][26]{Al}, \element[][53]{Mn}, and \element[][60]{Fe} as a function of time before present, roughly starting at the time when the Sun crosses the outer shock of the
LB, as derived from our fiducial simulation. The vertical dashed lines mark the explosion times of the last five SNe.}
         \label{im:ratiosplot}
   \end{figure}

\subsubsection{$\element[][60]{Fe}/\element[][26]{Al}$}
The $\element[][60]{Fe}/\element[][26]{Al}$ isotope ratios as ejected by the SNe fall within the range of 1--2  \citep[cf.][]{Limongi:2018}. However, the cyan line in Fig.~\ref{im:ratiosplot} illustrates that the SS encounters isotope ratios up to $\sim$30 during its residence in the LB shell (rightmost part of the plot). These high ratios reflect the combined radioisotopic abundances from the preceding SNe, which increase over time due to the differential radioactive decay of \element[][60]{Fe} and \element[][26]{Al}.
Each subsequent SN witnessed by the SS within the LHB yields \element[][60]{Fe}/\element[][26]{Al} ratios that closely resemble the values at the time of their ejection. Radioactive decay results in an additional increase in the ratio between each SN explosion, leading to a present-day value of about 3.9. 

Nucleosynthesis models generally predict a higher average Galactic flux ratio compared to the observed flux ratio \citep[e.g.][]{Sukhbold:2016,Feige:2018,Wang:2020}.
Therefore, the results obtained here, based on the yields of a specific nucleosynthesis model, are also inconsistent with gamma-ray observations and would differ when using other nucleosynthesis input data. Nonetheless, the comparison between our results and the observation is more complex, because our model only maps the local solar environment and omits the nucleosynthesis contribution of the most massive stars. The same should apply to the experimental ratios of $\element[][60]{Fe}/\element[][26]{Al}\ge 0.18$ determined from deep-sea sediments \citep{Feige:2018}, which are consistent with the results derived in our study.

\subsubsection{$\element[][53]{Mn}/\element[][60]{Fe}$}
The SN-ejected \element[][53]{Mn}/\element[][60]{Fe} isotope ratios strongly depend on the initial stellar masses, with a value of $\sim$0.3 for $\sim$20-M$_\sun$ stars, reaching the highest value of $\sim$30 for  $\sim$15-M$_\sun$ stars, and decreasing towards 1 for $\sim$13-M$_\sun$ stars. Therefore, as shown by the magenta line in Fig.~\ref{im:ratiosplot}, the SS is exposed to the combined radioisotopic abundances (including ratios ranging from 0.3 to 30) of the preceding SNe during its presence in the LB shell. The sharp increase only occurs as the SS enters the LHB region enriched by the preceding SN at $t\approx\unit[-4.48]{Myr}$, which injected a ratio of $\element[][53]{Mn}/\element[][60]{Fe} \approx 28.9$.
Each subsequent SN adds significantly less \element[][53]{Mn} compared to \element[][60]{Fe}, leading to a stepwise decrease of their ratio with time. The different decay rates slightly modify the ratio over time, resulting in a present-day interstellar ratio of about 3.7. This value is significantly lower than the measured present-day $\element[][60]{Fe}/\element[][53]{Mn}$ ratio from Antarctic snow samples, which falls in the range of 34--142 \citep{Koll:2019}. Assuming that our models are realistic, this may indicate that the nucleosynthesis ratios used here are too high by a factor of $\sim$10, which is still lower than the maximum they scatter around in the literature.

\subsection{Comparison to related work}
\label{sec:compmod}
To our knowledge, there is no work other than ours that has modelled the formation and structure of the LB from first principles constrained by the measured radioisotopic anomalies. What do exist, however, are studies that look at partial aspects of the problem.

\cite{Hyde:2018} re-determined the masses of potential SN candidates in UCL, LCC, and US, as well as Tuc-Hor. They confirmed UCL as the source of the recent $\element[][60]{Fe}$ peak and suggested Tuc-Hor as likely birth site of a SN progenitor producing the $\sim$7-Myr-old peak, although they stated that Tuc-Hor could be a candidate for either event. They excluded LCC and US as candidates for either of the peaks because they derived progenitor masses beyond their adopted SN mass range of 8--$\unit[18]{M_\odot}$ \citep{Smartt:2015}. The likely explosion sites were not determined. From follow-up calculations we can confirm that there may have been a single SN event in Tuc-Hor, in the mass range of 8--$\unit[15.9]{M_\odot}$ \citep[using isochrones taken from][]{Bressan:2012}. Stellar evolution models from \citet{Ekstroem:2012} predict corresponding lifetimes of 14--\unit[46]{Myr}. The age of Tuc-Hor was estimated to range between about \unit[30]{Myr} \citep{Torres:2008,Kraus:2014} and \unit[45]{Myr} \citep{Bell:2015}. In any case, the 8-M$_\odot$ star would not have yet exploded. If the explosion took place \unit[3]{Myr} ago, the mass of the exploding star would have been $\unit[10.8]{M_\odot}$ ($\unit[8.4]{M_\odot}$), applying a cluster age of \unit[30]{Myr} (\unit[45]{Myr}). Similarly, if the explosion occurred \unit[8]{Myr} ago, the corresponding mass would have been $\unit[11.7]{M_\odot}$ ($\unit[8.8]{M_\odot}$). Currently, Tuc-Hor is located close to the SS, with a centroid distance of $\sim$\unit[46]{pc}. Using \emph{Gaia} DR2 astrometry and the epicycle approximation, we traced back the centroid of Tuc-Hor in time. At $t=\unit[-3]{Myr}$, its distance to the SS was $\sim$\unit[80]{pc}, and at $t=\unit[-8]{Myr}$ it was $\sim$\unit[180]{pc} away. Hence, considering these distances, it is more likely that the SN contributed to the more recent $\element[][60]{Fe}$ signal, as it is unlikely for a single SN remnant located $\sim$$\unit[180]{pc}$ away, and then presumably still outside the LB volume, to have reached Earth.

\cite{Pelgrims:2020} derived the present-day shape of the LB shell purely on observational grounds, namely from 3D dust extinction maps of the LISM, including those of \citet[][see also Fig.~\ref{im:obsmaps}]{Lallement:2019}. They also expanded the inner surface of the shell in spherical harmonics up to a variable maximum multipole degree to make the complexity level of the modelled surface adaptable to a wide range of purposes, in their case to reconstruct the magnetic field in the LB shell from interstellar dust polarisation data at high Galactic latitudes. The LB shell they obtained roughly agrees with our results in terms of both size and thickness, and also shows no sign of a Galactic chimney.

\cite{Fujimoto:2020} turned the tables and, instead of reconstructing our solar neighbourhood in detail, addressed the question of how often certain conditions found in our local interstellar environment occur in an entire Milky Way-like galaxy. The three observational signatures they considered were: (1) the \element[][60]{Fe} influx onto the Earth's surface measured in deep-sea archives and Antarctic snow, (2) the broad distribution of \element[][26]{Al} observed in all-sky gamma-ray maps, and (3) the mean flux of diffuse soft X-ray emission. The data for this statistical study were snapshots of their own 3D galaxy-scale $N$-body + hydrodynamics simulation with an effective box size of (\unit[40.96]{kpc})$^3$ that was adaptively resolved down to $\unit[20]{pc}$. They found that stars in Sun-like orbits that meet the three constraints are rare but not exceptionally so ($\sim$2 per cent of their sample), and reside predominantly within kpc-scale SBs that form in the spiral arms. According to \citeauthor{Fujimoto:2020}, the residence time of a star there, and thus the duration over which it is exposed to elevated \element[][60]{Fe} fluxes, is given by the crossing time of the star across the spiral arm, being on the order of $10$--$\unit[100]{Myr}$, and is independent of the lifetime of the SB, which can clearly exceed the lower limit of this time range. These results are completely consistent with the picture we draw in this paper for the SS, which during its residence in the Local (or Orion) Arm may have passed not only through the LB but also through neighbouring SBs, in particular the Orion-Eridanus SB (see Sect.~\ref{sec:neighbsbs}).

\cite{Fry:2020} considered the survival and motion of charged dust grains in a magnetised SN remnant plasma. Applying a somewhat complicated and patched-up model in which grain dynamics, magnetic field, and one-dimensional (1D) hydrodynamics are treated separately, these authors claimed that a substantial fraction of iron grains is reflected back into the SN remnant interior by the swept-up magnetic field in the shell if they have a radius of about \unit[0.1]{\textmu m}, while larger grains have a high probability of being trapped, and smaller ones are destroyed by sputtering. However, these results have to be treated with caution because the authors neither performed self-consistent magnetohydrodynamic (MHD) simulations (in fact, the magnetic field was implemented ad hoc), nor is their 1D code suitable for treating a turbulent magnetic field. It was assumed that it would be sufficient to include the effects of RT instabilities, although this is not the only source of turbulence, given the huge amount of shear resulting from SN explosions. Another drawback is the assumption that one single explosion in Tuc-Hor would be responsible for the deposition of radioisotopes, while, as shown here, as many as 14 SNe could contribute to the \element[][60]{Fe} influx. Also, it was shown (e.g.~\citetalias{Schulreich:2017}; \citealt{Zucker:2022a}) that the LB is the result of multiple SN explosions, which, of course, has a severe influence on the structure of the SB density and magnetic field. We agree with \cite{Fry:2020} that magnetic fields may have an important influence on the dynamics of dust grains, but also on the acceleration and transport of Galactic CRs, both from within and outside the LB. Such an analysis will be the subject of future work.

\cite{Zucker:2022a} approximated the LB as a spherical shell expanding in a homogeneous background medium while cooling radiatively with a preset efficiency. For this idealised analytical model, which was developed on the basis of 1D numerical simulations, they determined the time of the first SN explosion, the time span between the explosions (which they assumed to be constant), the density of the ambient medium, and the thickness of the LB shell using Bayesian statistics, demanding that the formation of young stars within $\unit[200]{pc}$ of the Sun was triggered by the expansion of the supershell. The mean 3D positions of each parent star cluster required for this calculation were obtained by stellar tracebacks using \texttt{galpy}, which, as in our case, utilised \emph{Gaia} EDR3 as input. The explosion centre, assumed by them to be identical for all SNe, was set equidistant between UCL and LCC at the time of the first SN explosion. They subsequently derived the actual number of SNe that formed the LB using their modelling results and the present-day mass of the supershell estimated from local interstellar dust observations by calculating the ratio of the momentum of the supershell and the momentum contributed by a single SN on average (with $E_\mathrm{SN}=\unit[10^{51}]{erg}$). The number they obtained ($15\substack{+11 \\ -7}$) is consistent with ours ($14$), although their first SN explosion occurred $\unit[14.4\substack{+0.7 \\ -0.8}]{Myr}$ ago, about $\unit[4.2\substack{+0.7 \\ -0.8}]{Myr}$ before ours, within an ambient medium of density $\unit[2.7\substack{+1.6 \\ -1.0}]{cm^{-3}}$, which is higher than the maximum (i.e.~mid-plane) density in our fiducial simulation of $\unit[0.7]{cm^{-3}}$. The current radius they obtained for the LB of $\unit[165\pm 6]{pc}$ compares well with our final equivalence radius of $\unit[201]{pc}$, with \citeauthor{Zucker:2022a} neglecting the wind-driven phase and constraints from radioisotopic measurements. Their inability to reconcile their \texttt{galpy} back-calculations of UCL/LCC with our earlier ones based on the epicycle approximation \citep{Fuchs:2006, Breitschwerdt:2016} is only to a small extent due to this approximation, but rather to the fact that \citeauthor{Zucker:2022a} did not perform their calculations in the co-rotating but in a static LSR frame. As a result, their $X$ and $Y$ axes point only once per galactic year (in their case only at $t=0$) to the GC and in the direction of the Galactic rotation, respectively (with deviations increasing with the duration of their back-calculations). In our back-calculations, on the other hand, the alignment as required by the definition of the LSR is ensured at all times, which is reflected in the pronounced curvature of our stellar trajectories in the positive $X$ direction (compare Extended Data Figs.~4a and 4b of \citealt{Zucker:2022a}). Given the small peculiar motions of interstellar gas, we consider the coordinate frame we have chosen to be the most suitable for modelling the LISM in general and the LB in particular. Interestingly, the distance of the Earth to the centre of the first SN in our much more elaborate model ($\sim$$\unit[236]{pc}$) is hardly different from that derived by \citeauthor{Zucker:2022a} ($\sim$$\unit[300]{pc}$) and also the Sun enters the LB volume at a similar time (about $\unit[4.6]{Myr}$ ago for us and about $\unit[5]{Myr}$ ago for \citeauthor{Zucker:2022a}).

\cite{Chaikin:2022} performed 3D SPH simulations of singular, isolated SNe to study the propagation of \element[][60]{Fe} entrained in the gas and to numerically reconstruct the crust measurements of \cite{Wallner:2021}. By assuming their background medium to be at rest, homogeneous (in their fiducial runs with a density of $n_\mathrm{H}=\unit[0.1]{cm^{-3}}$ and temperature of $\unit[10^4]{K}$), and free of any radioisotopes, they ignored the observational fact that the SS is currently, and probably has been for several millions of years -- including the ages of the two measured \element[][60]{Fe} peaks -- in the highly turbulent, tenuous, and hot interior of at least one SB, radioisotopically contaminated by the previous sequential SN activity. Just like \cite{Fry:2020}, but interestingly without modelling decoupled dust dynamics, they found for their physically most complex setup with radiative cooling and a subgrid model for turbulent mixing that the ejecta and thus the \element[][60]{Fe}-enriched medium lag strongly behind the forward shock. Accordingly, they encountered the highest densities of \element[][60]{Fe} not in the shell of the SN remnant, but far behind it. They did not investigate whether this reverses to the situation shown by our simulations for the LB, where several SNe explode one after the other within a SN remnant, each of them sweeping up material from its predecessors\footnote{In the review by \cite{Diehl:2021}, there was an objection addressed to us that \cite{Krause:2018} had obtained a radioisotopic distribution (in their case of \element[][26]{Al}) for an SB that, qualitatively similar to the \element[][60]{Fe} distribution found by \cite{Chaikin:2022} for an isolated SN, showed high densities in the interior and low densities in the supershell. Although our simulations are based on the same grid code as those of \citeauthor{Krause:2018}, and also use passive scalars, we could only create such a situation for the LB if we switched off the radiative cooling completely so that the supershell is prevented from radiating away energy and thus from strongly condensing. We further point out that the line-of-sight averaged maps generated by \citeauthor{Krause:2018} are not directly comparable to our slice plots, which do not show averaged quantities.}. Instead, for reconstructing both \element[][60]{Fe} peaks observed, \citeauthor{Chaikin:2022} assumed the rather artificial scenario of two completely independent (i.e.~non-interacting) SN remnants expanding into the same uniform medium with a time separation of $\unit[3]{Myr}$. Like us, they assumed that the observer is not static. Unlike us, however, they did not determine a realistic solar trajectory via back-calculation in a local Galactic potential. In their model, the observer crosses the two SN blast waves almost perpendicularly on two linear trajectories with a constant speed of $\unit[30]{km\,s^{-1}}$, leading for \citeauthor{Chaikin:2022} to longer lasting \element[][60]{Fe}-flux pulses than in the static case. They further assumed the same \element[][60]{Fe} yield for both SNe ($\unit[10^{-4}]{M_\odot}$) and neglected radioactive decay during propagation in the ISM as well as gravitational forces.

\cite{Wehmeyer:2023} simulated a Galactic volume whose size ($\unit[2^3]{kpc^3}$) is between that of \cite{Fujimoto:2020} and ours, but still has the lowest spatial (\unit[50]{pc}) and temporal (\unit[1]{Myr}) resolution of all three. The simulations did not investigate the LB per se, but how $\element[][53]{Mn}$, $\element[][60]{Fe}$, $\element[][182]{Hf}$, and $\element[][244]{Pu}$ with Type Ia SNe, CC SNe, intermediate-mass stars, and KNe as assumed exclusive sources, respectively, propagate through the general ISM. Despite their different origins, the increases of the radioisotopic densities often coincide, which these authors attributed to CC SNe being the most dominant propagation mechanism. This finding is basically compatible with our results for the propagation of CC SN radioisotopes and $\element[][244]{Pu}$ from KNe, given the much lower time resolution of \citeauthor{Wehmeyer:2023}. In contrast to their model, we considered $\element[][53]{Mn}$ to be produced by CC SNe -- the only SN type in our simulations, motivated by the rate of Type Ia SNe in the Milky Way being lower than that of CC SNe by a factor of 10 \citep{Matteucci:2006} -- with yields similar to or higher than the constant $\unit[10^{-4}]{M_\odot}$ applied by \citeauthor{Wehmeyer:2023} (see Table \ref{tab:input}).

\cite{Ertel:2023} fitted the $\sim$3-Myr-old \element[][60]{Fe} peak in the deep-sea sediment data from \cite{Ludwig:2016} and \cite{Wallner:2016} with various simple functions (cut exponential, Gaussian, Lorentzian, `sawtooth', `reverse sawtooth', symmetric triangle, asymmetric triangle or `sharktooth'). They found that the data do not specifically favour any of these fits (whose physical motivation is partly unclear to us), but when all the data are combined, the timescale for the deposition of \element[][60]{Fe} is $>$\unit[1.6]{Myr}, and thus longer than the timescale for the passage of a single Sedov-like SN blast ($\lesssim$\unit[0.1]{Myr}). \citeauthor{Ertel:2023} interpreted this as evidence for the validity of the model of \cite{Fry:2020} (i.e.~that the dynamics of the \element[][60]{Fe}-enriched dust is separate from but coupled to the evolution of the blast plasma). However, they also admitted that the data do not exclude a multi-SN scenario as proposed by us.

\section{Discussion}
\label{sec:discussion}
\subsection{Imperative improvements}
In our previous works \citep[e.g.][]{Fuchs:2006,Breitschwerdt:2016} we made some assumptions that \citet{Neuhaeuser:2020} pointed out to be problematic. Hence, we adjusted our model accordingly by (1) replacing the IMF of \citeauthor{Massey:1995} (\citeyear{Massey:1995}; $\alpha=2.1$) with that of \citeauthor{Kroupa:2001} (\citeyear{Kroupa:2001}; $\alpha=2.3$) yielding a lower amount of stellar explosions, (2) using a lower cluster age of \unit[20]{Myr} for UCL/LCC instead of \unit[22.5]{Myr}, and (3) deriving the stellar lifetimes on the basis of the rotating stellar evolution models of \citet{Ekstroem:2012} instead of applying the \citet{Schaller:1992} isochrone fit from \citet{Fuchs:2006}, which is actually only valid above \unit[12]{M$_\odot$}.

For the scope of our simulations, we had to combine different stellar evolution calculations (in our case \cite{Ekstroem:2012} and \cite{Limongi:2018}), which always bears the potential of inconsistencies between the models \citep[e.g.~nucleosynthesis yields; see][]{Lawson:2022}. This is also reflected by the difference of 1--2 orders of magnitudes for the total wind-released \element[][26]{Al} mass for stellar masses below about $\unit[15]{M_\sun}$. However, the higher pre-SN yields from \citeauthor{Limongi:2018} are still about one order of magnitude below their explosive yields and might therefore be safely neglected if we had used them in our numerical simulations. Our choice is at least consistent in that only models for solar metallicities that take stellar rotation into account were used.

Our newly found lower SN candidate mass limit for UCL/LCC of \unit[12.9]{M$_\odot$} is higher than in our previous studies (\unit[8.8]{M$_\odot$}), now in agreement with \citet{Hyde:2018} and \citet{Neuhaeuser:2020}, and excluding EC SNe as prime candidates. Furthermore, we now calculated the stellar trajectories by means of test-particle simulations using a realistic Milky Way potential \citep{Barros:2016} instead of the epicycle approximation. Our comparative studies between the two methods showed differences of more than \unit[20]{pc} after \unit[10]{Myr} implying that the here adopted approach is indeed  necessary for the considered evolution times.

A potentially even greater influence on the stellar trajectories has the value for the peculiar motion of the Sun, which is still particularly uncertain along the $Y$ direction (i.e.~along Galactic rotation). Here, higher values of the corresponding velocity component, $V_\odot$, can lead to a `compression' of the trajectory of nearby star clusters in the $Y$ direction, and thus to an overall shorter distance travelled in the LSR frame. Our previously used value, $V_\odot = \unit[5.2]{km\,s^{-1}}$ \citep{Dehnen:1998}, led to the somewhat paradoxical situation that the bulk of the progenitor population UCL/LCC seemingly entered today's LB volume only shortly before $t\approx\unit[-5]{Myr}$ \citep[see fig.~6 of][]{Fuchs:2006}, although SNe should have occurred in the population well before that. Also the explosions were considerably off-centre with respect to the LB. By using the more modern value of \cite{Schoenrich:2010} of $V_\odot =\unit[12.24]{km\,s^{-1}}$ in this work, we cleared up this paradox criticised by \cite{Zucker:2022a}. As in \citeauthor{Zucker:2022a}, who used a slightly higher but older value ($V_\odot = \unit[15.4]{km\,s^{-1}}$; \citealt{Kerr:1986}) for their calculations, the trajectories of the progenitor populations (UCL/LCC and V1062~Sco) now completely lie in the interior of the present-day LB volume (as verifiable from the trajectories of their perished members in the last column of Figs.~\ref{im:ts-sy}, \ref{im:ts-sz}, and \ref{im:ts_sx}), and the explosion sites of almost all SNe are now not far from the present-day LB centre. As noted by \citeauthor{Zucker:2022a}, this is achievable for the fairly wide interval of $V_\odot=10$--$\unit[16]{km\,s^{-1}}$, and thus for most of the values of $V_\odot$ available in the literature (see e.g.~\citealt{Ding:2019} for a relatively recent listing).

\subsection{Caveats and limitations}
\label{sec:caveats}
Among the physical processes and conditions that are not (yet) taken into account in our numerical simulations are heat conduction, ionising radiation from the stars, NEI effects, and interstellar magnetic fields. 

The effect of heat conduction would be to increase the density and lower and homogenise the temperature in the interior of both the wind-blown bubbles \citep{Lancaster:2021} and the later SB \citep{Badry:2019}, as thermal energy would be transferred from the hot cavities to the cooled shells and parts of the shells would be evaporated by the heat flow \citep{Tomisaka:1986}. This would likely provide better agreement with maps of the diffuse soft X-ray background \citep{Snowden:1997}, whose $\sim$$\unit[0.25]{keV}$ component is expected to primarily originate from the thermal emission of the hot interior of the LB \citep{Galeazzi:2014}. The relatively high temperature ($\sim$$\unit[10^7]{K}$) we obtained for the LHB gas is not perfectly consistent with this expectation\footnote{Another, and more direct, way of lowering the temperature inside the hot bubble would be mass loading of the flow \citep{Hartquist:1986, Dyson:1987}, for example by the LB shock overrunning dense clumps in an ambient medium with inhomogeneities.}. And although it would not matter much for the present state of the LB, since any structures generated during the wind-driven phase get more or less `erased' shortly after the onset of the SN-driven phase, heat conduction would also alter the bow shocks of the LB progenitor stars \citep[see e.g.][]{Meyer:2014}.

As recently shown by \cite{Dwarkadas:2022} in 2D simulations of isolated wind-blown bubbles, ionising photons from the stars would modify the hydrodynamical structure of the bubbles as well, introducing ionised regions of higher number densities. In addition, emerging ionisation front instabilities would further excite the turbulence in the wind-blown cavities, favour the formation of dense clumps and filaments, and make deviations from spherical symmetry even stronger. However, with the successive deaths of the massive stars during the SN-driven phase, the influence of photoionisation should quickly diminish greatly.

Plasma heating by successive SN explosions and cooling by fast adiabatic expansion of the LB are typical examples for an underionised and overionised plasma, respectively. The LB plasma is therefore in a general state of NEI, as both shock heating and adiabatic cooling occur on timescales much shorter than ionisation and recombination. Simulating a thermally and dynamically self-consistent evolution of the plasma offers the unique possibility of tracing the ionisation history. This allows one to derive the evolution timescale of the LB by calculating ion line ratios of \ion{C}{IV}, \ion{N}{V}, and \ion{O}{VI}, and by comparison with ultraviolet observations, such as FUSE data, to infer when the last reheating of the LB by SNe has taken place. For a detailed discussion, see \cite{Avillez:2012}, who have deduced a time window of 0.5--$\unit[0.8]{Myr}$ after the last SN has occurred.

Magnetic fields would add to the external pressure and tension that the interstellar bubbles have to compete against, favouring their expansion along the mean-field direction. Since field lines are swept up by the shells and comparatively high field strengths are thus built up there, they would mitigate shell instabilities and thereby impede turbulent mixing, possibly also reducing the amount of $\element[][244]{Pu}$ arriving in the LB cavity. However, the local magnetic field in the Galactic disc should generally be too weak (\unit[1.6]{\textmu G}; \citealt{Xu:2019}) to significantly influence the dynamical evolution of the LB, especially after the onset of the SN-driven phase. This should apply even more to the SN-mediated fluxes of radioisotopes to which the SS is exposed, since the medium in which these shock waves propagate -- the interior of the LB -- should be almost field-free \cite[cf.][]{Avillez:2005}.

One point noted by \cite{Neuhaeuser:2020} that we still neglected in the present study is the possible presence of binaries (and thus mass transfer between companion stars). This would modify the IMF and thus the predicted number, masses, and lifetimes of the LB progenitors, and, as a consequence, to some extent, their explosion times and sites.

\subsection{Neighbouring superbubbles}
\label{sec:neighbsbs}
In contrast to our previous works, we have not considered the evolution of the LB together with that of an immediately neighbouring SB lying towards the GC region, which is commonly seen as the explanation for Loop~I (LI), the angularly largest and brightest of the radio-continuum loops stretching across the northern Galactic sky. The interior of LI emits soft X-rays and its eastern base is an elongated structure of enhanced brightness seeming to rise from the Galactic plane, known as the North Polar Spur (NPS). The decision to ignore this potential `boundary condition' this time was made on the basis that its presence should not play a major role in the generation of the measured radioisotopic anomalies, and because there is currently no coherent model for LI/NPS that explains all observational evidence. Thus, in the last few years, the voices of those who do not see LI/NPS as a local source but instead locate it in the GC region grew louder. The numerous pros and cons for these two hypotheses were compiled by \cite{Lallement:2022} in a recent review article. What is widely agreed upon is that the centre of LI/NPS is further away than originally thought ($\sim$$100$--$\unit[200]{pc}$), very likely more than $\unit[700]{pc}$. Although X-ray and synchrotron emission observations could be reconciled with a SB at this distance, whose closest part at high latitudes reaches up to $100$--$\unit[200]{pc}$ from the Sun and thus in principle allows direct interaction with the shell of the LB, which \cite{Egger:1995} claimed to have observed, the locations of O and B stars and the 3D distribution of interstellar dust clouds argue against it, especially if the SB were to extend over both sides of the Galactic plane. On the other hand, the long-distance hypothesis, which envisages the NPS as a gigantic extension of the northern Fermi bubble known to be blown by the GC, and LI representing the bounding shock front, suffers from being incompatible with the estimated proximity of the synchrotron source at high latitude in the north and the observed absence of loops in the west and south, which would however be expected from a giant synchrotron-emitting shell \citep[see][and references therein]{Lallement:2022}. Another argument for the proximity of LI is the natural explanation for the origin of the local clouds as a product of a hydromagnetic RT instability operating in the interaction zone with the LB \citep{Breitschwerdt:2000}.

But even leaving LI/NPS aside, the LB is still flanked by a number of other SBs, including GSH~238+00+09 \citep{Heiles:1998,Puspitarini:2014}, the Perseus-Taurus SB \citep{Bialy:2021}, and the Orion-Eridanus SB. The latter has a $\unit[200]{pc}$ wide and $\unit[250]{pc}$ long cavity enclosed by a shell of neutral gas expanding at $\unit[20]{km\,s^{-1}}$, with the part facing the LB extending to a distance of $150$--$\unit[200]{pc}$ from the Sun \citep{Joubaud:2019}. Its formation is attributed to the ionising radiation and energetic winds of tens of massive stars, as well as a series of 10--20 SNe \citep{Bally:2008} that occurred at a rate of about $\unit[1]{Myr^{-1}}$ over the past $\unit[12]{Myr}$ \citep{Voss:2010}. The composite structure of the SB likely evolved from near to far along the blue stream of massive stars, identified by \cite{Pellizza:2005} and \cite{Bouy:2015} in front of the Orion clouds, over a length of $\unit[150]{pc}$ along the major axis of the SB. The ages of the stars in the stream range from $\unit[20]{Myr}$ (near) to $\unit[1]{Myr}$ (far) \citep{Joubaud:2019}. The closest and most active of the Orion molecular clouds, Orion~A, presumably witnessed a major SN  event (`Orion Big Blast') about $6$--$\unit[7]{Myr}$ ago \citep{Kounkel:2020,Grossschedl:2021}. This is particularly interesting in that the Sun may have passed through the Orion-Eridanus SB before entering the LB and thus may have been exposed to this and/or other SNe. The rough coincidence in time with the older peak $6.5$--$\unit[8.7]{Myr}$ ago, measured for $\element[][60]{Fe}$ and apparently also included in the $\element[][53]{Mn}$ data, is in any case remarkable. The width of the peak ($\Delta t\approx\unit[2.2]{Myr}$) would in principle be consistent with this scenario, since the Sun had a mean peculiar velocity of $\overline{|\vec{V}_\odot|}\approx\unit[18.0]{km\,s^{-1}}$ during this time interval (according to our back-calculation), and was thus exposed to elevated $\element[][60]{Fe}$ and $\element[][53]{Mn}$ fluxes over a distance of $\overline{|\vec{V}_\odot|}\,\Delta t\approx\unit[40.5]{pc}$, which is below the present-day extent of the Orion-Eridanus SB. Similarly, the Sun's residence in a dilute environment, presumably cleared of $\element[][244]{Pu}$, such as the interior of a SB, would conclusively explain why, contrary to our simulations, the measured $\element[][244]{Pu}/\element[][60]{Fe}$ ratio was significantly reduced in the time range $4.57$--$\unit[9.01]{Myr}$ ago. We plan to quantitatively test the validity of this scenario in a forthcoming paper.

\subsection{Star formation at the Local Bubble boundary}
   \begin{figure*}
   \resizebox{\hsize}{!}
            {\includegraphics{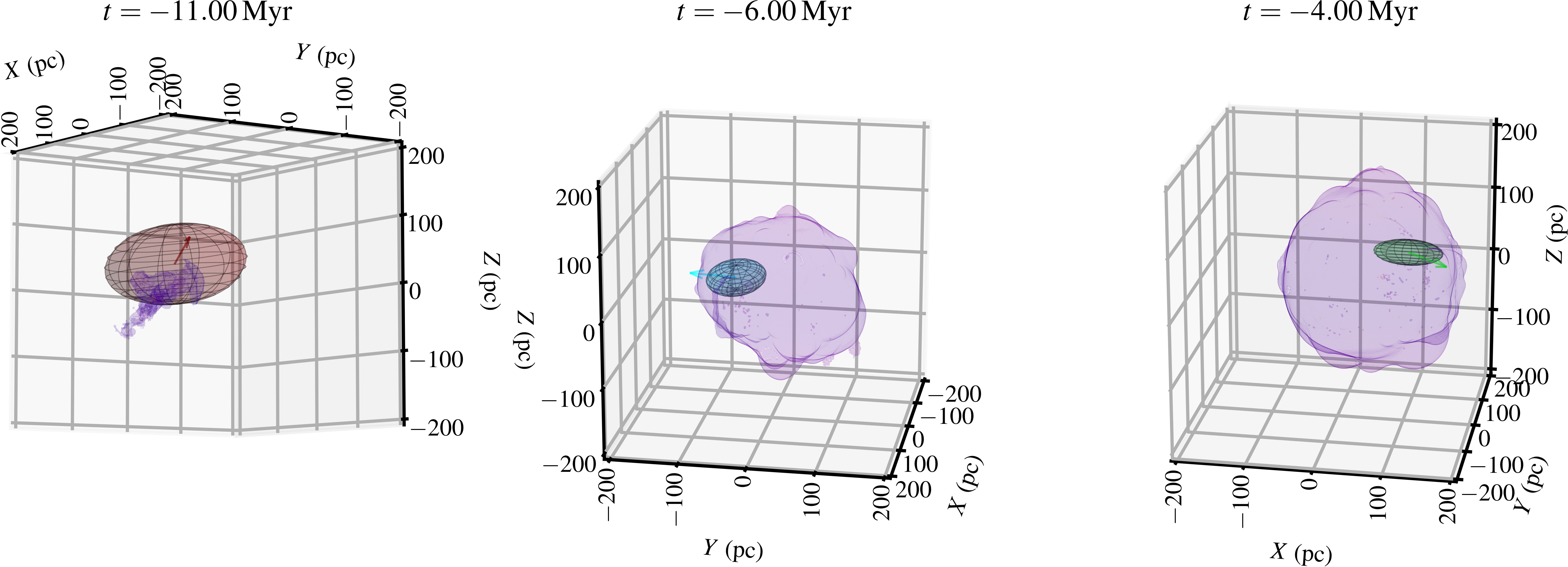}}
      \caption{Presumed birthplaces of Sco-Cen subgroups with respect to the evolving LB. Shown in purple in each panel is the semi-transparent enveloping surface of that volume in our fiducial simulation where the gas speed exceeds $\unit[1]{km\,s^{-1}}$, being representative of the medium belonging to the LB, together with the normalised velocity vectors of the centroids (arrows) and the centroids' 3$\sigma$ error ellipsoids computed for the 10\,000 realisations of the populations US (\emph{left panel}), Lupus (\emph{middle panel}), and Ophiuchus (\emph{right panel}) at their respective birth times (numbers at the top). The coordinates of the centroid $(X_\mathrm{c},Y_\mathrm{c},Z_\mathrm{c})$ and its peculiar velocity components $(U_\mathrm{c},V_\mathrm{c},W_\mathrm{c})$ are $(73.3\pm 51.6,4.6\pm 25.6,39.3\pm 20.1)$\,pc and $(3.9\pm 4.8,-5.8\pm 2.4,5.4\pm 1.3)$\,km\,s$^{-1}$ for US, 
      $(90.9\pm 23.3,-35.3\pm 14.1,40.6\pm 6.2)$\,pc and $(6.6\pm 3.9,-7.4\pm 2.2,2.5\pm 1.3)$\,km\,s$^{-1}$ for Lupus, and $(107.8\pm 17.1,-15.9\pm 4.9,65.4\pm 6.5)$\,pc and $(4.8\pm 4.1,-3.7\pm 1.2,-0.2\pm 1.5)$\,km\,s$^{-1}$ for Ophiuchus, with errors indicating 1 standard deviation.}
         \label{im:birthplaces}
   \end{figure*}
We can confirm, at least qualitatively, the main hypothesis of \cite{Zucker:2022a}, namely that star formation near the Sun was triggered by the expansion of the LB, on the basis of our calculations for all Sco-Cen subgroups younger than the progenitor populations UCL/LCC and V1062~Sco. As shown in Fig.~\ref{im:birthplaces}, the birthplaces of US (dark red ellipsoid in the left panel), Lupus (cyan ellipsoid in the middle panel), and Ophiuchus (green ellipsoid in the right panel) were overrun by shock waves associated with the evolution of the LB (sharp boundaries of the purple volume in each panel) shortly before the indicated birth times, with the mean space velocity vectors of the populations (arrows) always roughly coinciding with the closest shock normals (being in agreement with \citeauthor{Zucker:2022a}). For US, whose birth time is before the onset of the SN-driven phase, it is the stellar wind bow shocks of the LB progenitor stars that may have compressed one or more molecular clouds located in this region, rendering them gravitationally unstable, whereas for the two younger populations, Lupus and Ophiuchus, it was probably the expanding supershell. We say `probably' because this statement (just like that of \citeauthor{Zucker:2022a}, by the way) is only based on how the trajectories of the stellar populations lie relative to the evolving LB boundary. A true feasibility study can only be carried out using hydrodynamic simulations that also include the star formation itself, which we defer to a forthcoming paper.

\subsection{Signal timings and durations}
\label{sec:stimings}
The coarse time resolution of the $\element[][244]{Pu}$ measurements leaves uncertainty regarding its arrival on Earth in relation to $\element[][60]{Fe}$. In our model, we assumed a homogeneous enrichment of $\element[][244]{Pu}$ in the LISM from rare r-process events. Consequently, if the $\sim$7-Myr-old $\element[][60]{Fe}$ peak was indeed created by the SS passing through another SB, we would expect an anti-correlation between $\element[][244]{Pu}$ and $\element[][60]{Fe}$. Specifically, during the residence in the SBs, coinciding with the $\element[][60]{Fe}$ signals, we would anticipate minimal $\element[][244]{Pu}$. Outside of the SBs, we would expect an increased influx, reaching its maximum shortly ($\sim$\unit[60]{kyr}) before the arrival of SN-derived $\element[][60]{Fe}$, originating from material swept up by the supershells. However, if future higher-resolution measurements demonstrate the simultaneous occurrence of $\element[][244]{Pu}$ and $\element[][60]{Fe}$, it could indicate either the production of $\element[][244]{Pu}$ in CC SNe after all or the occurrence of a rare type of SN within the LB itself, as also envisaged by \cite{Wang:2021}.

Regarding the prolonged duration of the $\element[][60]{Fe}$ signal, various explanations have been proposed. For instance, alternative transport mechanisms have been suggested, such as significant decoupling between gas and dust dynamics \citep{Fry:2020}, or post-depositional processes like particle diffusion \citep{Wallner:2021}. In our previous studies, we proposed the LB shell as the source of the $\sim$3-Myr-old $\element[][60]{Fe}$ signal \citep[e.g.][]{Breitschwerdt:2016}. However, we now discard this scenario since if the SS had entered the LB only $\sim$\unit[3]{Myr} ago, the time required to reach its current position near the centre would have been insufficient.

Furthermore, in the present study, we reproduced the broad $\sim$3-Myr-old $\element[][60]{Fe}$ peak by considering multiple sequential SN events, without explicitly accounting for dust dynamics or post-depositional diffusion, which could potentially influence the distribution of the SN signals. The broadening of the signals is solely due to the low temporal resolution of the deep-sea FeMn crusts, which demonstrates that it is highly unlikely to detect individual SN events within these records. This underscores the importance of investigating geological archives with improved time resolution and reduced particle migration to detect the individual SN signals within each currently known $\element[][60]{Fe}$ peak.

\section{Summary}
\label{sec:summary}
In this paper, we presented 3D hydrodynamic simulations with subparsec resolution of the formation and evolution of the LB, as well as the associated transport of radioisotopes to Earth. The simulations are based on the model used in \citetalias{Schulreich:2017} to determine the number, timing, and locations of the near-Earth CC SN explosions, which we have now updated and improved in many ways.

The basis for the present work was the sample of members of the Sco-Cen complex, which has already emerged in the past as the most likely source of most if not all of these SNe, compiled by \cite{Luhman:2022}. However, our previous calculations had relied on data from the \textsc{Hipparcos} satellite and not yet on data from \emph{Gaia}, which is superior in terms of sensitivity, astrometric precision, and sky coverage. This has now been remedied, as the selection of \citeauthor{Luhman:2022} is based on \emph{Gaia} EDR3. We calculated the number of missing (i.e.~already exploded) stars by fitting the more recent IMF of \cite{Kroupa:2001} instead of the IMF of \cite{Massey:1995} and thus obtained a total of 14 explosions, 13 of which occurred in UCL/LCC and one in V1062~Sco -- both being populations of Sco-Cen. The timing of these explosions is also related to the initial masses of the missing massive stars, as these dictate their lifetimes. To determine the latter, we linearly interpolated between the modern rotating stellar evolution tracks for solar metallicity derived by \cite{Ekstroem:2012}. Assuming that all stars in a population are born at the same time, we obtained the explosion times as differences between the lifetimes and the population ages, which are also given in \citeauthor{Luhman:2022}'s paper. As initial masses of the SN progenitor stars, we chose those values for which the IMF reaches its mean value in the intervals that we have determined to contain exactly one star, corresponding to the distribution with the highest probability. We furthermore enhanced the derivation of the hypothetical (or pseudo-) trajectories of the SN progenitors, incorporating a new Monte Carlo-type approach that utilised temporal back-calculations of numerous realisations of the Sco-Cen populations to which the progenitors belonged. In contrast to our earlier work, these back-calculations were no longer based on the epicycle approximation but were full-blown test-particle simulations using a realistic Milky Way potential \citep{Barros:2016} and a contemporary value for the peculiar velocity of the Sun \citep{Schoenrich:2010}, which we also traced back in time. As explosion sites and thus endpoints of the trajectories of the SN progenitors, we chose the maxima of the 6D phase-space PDFs at the times of the explosions.

The hydrodynamic evolution of the LB, which, compared to \citetalias{Schulreich:2017}, also includes the effects of the age- and initial mass-dependent stellar winds blown by the SN progenitors, as prescribed by the \citet{Ekstroem:2012} tracks, was studied in an inhomogeneous LISM set up to be in hydrostatic equilibrium with the aforementioned gravitational potential. We also extended our simulations to include additional radioisotopes alongside $\element[][60]{Fe}$, namely $\element[][26]{Al}$, $\element[][53]{Mn}$, and $\element[][244]{Pu}$, where the stellar wind-derived $\element[][26]{Al}$ yields and the explosive yields of $\element[][26]{Al}$, $\element[][53]{Mn}$, and $\element[][60]{Fe}$ are based on the rotating models of \cite{Ekstroem:2012} and \cite{Limongi:2018}, respectively.

With all these improvements, we are now converging to a comprehensive picture of the origin of the LB.
\begin{enumerate}
    \item Interpreting the temporal distribution of the radioisotopes found through accelerator mass spectrometry in deep-sea FeMn crusts, nodules, and sediments \citep[see e.g.][]{Wallner:2016} coherently, particularly $\element[][60]{Fe}$ (which is almost exclusively due to stellar nucleosynthesis), we conclude that the LB must be the result of a series of SN explosions.
    
    \item The location of the recent $\element[][60]{Fe}$ peak about 2--\unit[3]{Myr} ago naturally coincides with the closest approach of the progenitors' trajectories with respect to the SS, providing strong evidence that members of the Sco-Cen populations UCL/LCC and V1062 Sco are responsible for the origin of the LB. These trajectories have now been extracted from a Monte Carlo-based ensemble of simulated pseudo-trajectories, and thus are not restricted by any a priori assumptions but solely constrained by the observational errors in the \emph{Gaia} data.

    \item The satisfactory explanation, or at least consistency of other radioisotopic data, such as $\element[][26]{Al}$, $\element[][53]{Mn}$, and $\element[][244]{Pu}$ with the same model, adds to its confidence, which will be steadily improved as soon as more precise data are available.   

    \item The recent influx of $\element[][60]{Fe}$ discovered in Antarctic snow and deep-sea sediments  \citep{Koll:2019,Wallner:2020} seems to argue against a SN origin since the last SN in our sample occurred \unit[0.88]{Myr} ago (see Table~\ref{tab:input}). Therefore, the passage of the SS through the LIC (thus collecting $\element[][60]{Fe}$-carrying dust particles) has been invoked as a possible source. This hypothesis could actually be tested if data were available reaching back to the time when then Sun was still outside, then manifesting itself in a significant drop of $\element[][60]{Fe}$ flux. However, we believe from our simulations that the still ongoing influx is due to the turbulence excited by the most recent SNe, with some contribution from reflected shocks as they hit the contact discontinuity in the LB shell.

    \item For the first time, also the ejections of radioisotopes by stellar winds have been included, predominantly affecting $\element[][26]{Al}$. Moreover, the expansion of the LB into a pre-SN structured stellar wind cavity has been taken into account.
    
    \item Since all 14 explosions that we derived from the most likely IMF mass distribution originate from star clusters, all but one even from the same one, it is highly unlikely that the LB was produced by a single SN event, as has been claimed in the past \citep[e.g.][]{Fry:2020}.

    \item We confirm the possibility of triggered star formation in molecular clouds \citep[see][]{Zucker:2022a} from our numerical investigations. The birth of nearby clusters younger than the LB, such as Lupus or Ophiuchus, could have been initiated by the forward shock of the expanding LB, overrunning molecular clouds just before the isochronal age of the clusters.
    
    \item According to the peculiar solar motion data by \cite{Schoenrich:2010}, the SS was born \emph{outside} the LB, and entered it only \unit[4.6]{Myr} ago, meaning that the older and smaller peak in the $\element[][60]{Fe}$ data around 6--\unit[9]{Myr} ago does not derive from SN explosions within the LB, but must come from a different SB that the SS has crossed before. This will be the subject of a forthcoming paper. 

    \item The transport of the radioisotopes is most probably due to dust particles, which are naturally generated in the post-SN expanding plasma and allow the isotopes to overcome the solar wind ram pressure. However, how these grains move through the LB is not clear. We expect their size to be comparatively small due to their heavy sputtering in SN shocks, especially in the reverse shock. Therefore, we assumed them to behave like a passive scalar. However, further 3D hydrodynamic and MHD simulations are necessary to clarify this. 
    
    \item The more recent entry of the SS into the LB also implies that the peak at 2--\unit[3]{Myr} before present cannot be due to the dust-incorporated radioisotopes swept up in the LB shell but should be due to individual explosions. Therefore, one would expect more peaks, like in Fig.~\ref{im:fluxplot}, around this time in the data, if their time resolution is high enough. This is a prediction that could be tested. However, it should be mentioned that sedimentation on the ocean floor is accompanied by some diffusion across the various layers contaminated by the explosions, resulting in the signal being smeared out. Therefore, probably only a resolution higher than the one achieved so far has the potential to resolve the issue.  
\end{enumerate}

The LB is just one of several SBs in the solar neighbourhood. Therefore, it serves as an ideal test laboratory for studying the evolution of the ISM in star-forming regions.

\begin{acknowledgements}
We are particularly indebted to Kevin Luhman for helpful correspondences and for sharing with us supplementary data from his Sco-Cen analysis. We thank Maurice K\"{u}nicke for his extensive preliminary work with the \texttt{galpy} package, Victoria Herpel for her preliminary kinematic and photometric analysis of nearby young stellar associations using \emph{Gaia} DR2 to identify past SNe, and Emre Elmal{\i} for assisting in the preparation of the test-particle simulations.  We are grateful to Thomas Faestermann, Gunther Korschinek, and Toni Wallner for answering our questions about their data. We acknowledge enlightening conversations with Jo\~{a}o Alves, Adrienne Ertel, Josefa Gro{\ss}schedl, Hendrik Heinl, Dominik Koll, Jonathan Mackey, Britton Smith, Allard Jan van Marle, and Catherine Zucker.

This work has made use of data from the European Space Agency (ESA) mission \emph{Gaia} (\url{https://www.cosmos.esa.int/gaia}), processed by the \emph{Gaia} Data Processing and Analysis Consortium (DPAC, \url{https://www.cosmos.esa.int/web/gaia/dpac/consortium}). Funding for the DPAC has been provided by national institutions, in particular the institutions participating in the \emph{Gaia} Multilateral Agreement. 

This publication has made use of data products from the Two Micron All Sky Survey, which is a joint project of the University of Massachusetts and the Infrared Processing and Analysis Center/California Institute of Technology, funded by the National Aeronautics and Space Administration and the National Science Foundation.

This research made use of \texttt{yt}, a toolkit for analysing and visualising quantitative data \citep{Turk:2011}.

J.F. acknowledges funding by the European Union (ERC, NoSHADE, 101077668). Views and opinions expressed are however those of the author(s) only and do not necessarily reflect those of the European Union or the European Research Council Executive Agency. Neither the European Union nor the granting authority can be held responsible for them.

Author contributions: M.M.S.~thought up the extensions and improvements of the original LB model invented by D.B., performed and analysed all numerical simulations, produced all figures, curated the research data, and wrote the initial draft of the paper. J.F.~contributed to the interpretation of the results and to the writing of Sects.~\ref{sec:intro}, \ref{sec:compmeas}, \ref{sec:compmod}, and \ref{sec:discussion}. D.B.~contributed to the writing of Sects.~\ref{sec:compmod}, \ref{sec:caveats}, and \ref{sec:summary}, and wrote Appendix \ref{app:A}.
\end{acknowledgements}

\bibliographystyle{aa}
\bibliography{LBII}

\begin{appendix}
\section{Treating supernova-generated local interstellar dust as a passive scalar}
\label{app:A}
Although counter-intuitive, dust formation in CC SN explosions is a reality \citep[see][and references therein]{Brooker:2022}. On the one hand, SN ejecta contain a large fraction of chemically enriched material necessary for dust formation. On the other hand, the SN reverse and forward shocks are efficient destroyers of dust grains. Physically, however, a suitable window in density and temperature arises when the forward shock breaks out, causing the temperature to decrease appreciably in the expanding medium. Within this recombining plasma, gas-phase chemical reactions are initiated, altering the composition just before the nucleation once the temperature drops below about \unit[5000]{K}. All this occurs on short timescales, which vary for different species and happen well before the reverse shock detaches from the contact discontinuity. Hence, any dust grains that have been decoupled and left behind will be strongly eroded, with the surviving ones being subject to strong turbulent transport (see below). The grains that remain in the post-shock region, and are therefore small enough, will most likely experience Epstein drag since their size is small compared to the mean free path in a dilute hot gas, and thus their stopping times could be quite large \citep[see e.g.][]{Fujimoto:2020}.

Clearly, the dust dynamics in general, and behind SN shocks in particular, is quite complicated since many processes on various scales are present (radiative forces, drag forces, sputtering and shattering, electric and magnetic forces as the grains are charged, etc.), and even interact with each other. Therefore, it is not surprising that grains start to drift through the plasma. It has been shown in a series of papers, however, that the drift motion itself is the source of resonant instabilities both in hydrodynamic and MHD cases \citep{Squire:2018, Hopkins:2018}, if the drift velocity times the wave vector equals the gyrofrequency, somehow reminiscent of the streaming instability of CRs through a magnetised plasma. This causes the grains to bunch and clump, essentially decreasing their drift speed. We consider all these intricacies to be well beyond the scope of this paper and, for simplicity, stick to the passive scalar assumption.

When dust however decouples from the gas, then it is certainly subject to the strong turbulence behind the SN shocks.
We thus picture the particles to be immersed in an ensemble of eddies, interacting with their self-advected velocity field. Their motion is ballistic within a correlation length $l$ and diffusive in regions where large gradients occur, hence on smaller scales. The advection-diffusion equation for a passive scalar of concentration $C$ is given by
\begin{equation}
    \frac{\partial C}{\partial t} + (\vec{u}\cdot \vec{\nabla}) C = \alpha_\mathrm{d}\,\nabla^2 C
\end{equation}
where $\alpha_\mathrm{d}$ is the diffusion coefficient and $\vec{u}$ 
is the bulk fluid motion. Although the LB is expanding, we assume for the following arguments that the bulk motion of the fluid is at rest and characterised by strong turbulence, which is homogeneous and steady in a statistical sense.
 
Now, from a physical perspective, a dust particle will travel an average distance of $\bar{\vec{x}} = \vec{u}\,\tau$, when immersed in an eddy within a turn-over time $\tau$, roughly corresponding to its mean free path, while on crossing eddies, $\langle |{\vec{x}}|\rangle = \sqrt{2\,\alpha_\mathrm{d}\,t}$, corresponding to a standard deviation, if turbulent diffusion were a random Gaussian process, which it is not, as we shall show; time averaging has been replaced by the ensemble average $\langle\dots\rangle$.  

The assumption of eddy diffusion being Markovian is generally not true, as the eddies imply a hierarchy of length scales with a non-Gaussian distribution. Let us consider the motion of a dust particle by evaluating its root mean square (RMS) distance $x_\mathrm{rms} = \sqrt{\langle x^2\rangle}$ from time $t=0$ to time $t>0$, where the number of crossing eddies is sufficiently large to guarantee a trajectory that is determined by a stochastic process, meaning that the particle changes its direction frequently enough. Since $\mathrm{d}\vec{x}/\mathrm{d}t = \vec{u}_\mathrm{L}$ we have $\vec{x} = \int_0^t \vec{u}_\mathrm{L}(t^\prime)\, \mathrm{d}t^\prime$, where $\vec{u}_\mathrm{L}$ is the Lagrangian velocity of the particle. We need to distinguish between the correlation time of the (large) eddies in the turbulent flow, which is denoted by $\tau$, and the timescale over which the dust particle retains memory of its trajectory from when it started at $t=0$, known as the Lagrangian correlation time $t_\mathrm{L}$. In incompressible turbulence, the scaling after Kolmogorov of the eddy turn-over timescale for an invariant energy dissipation rate is given by $\tau \sim l/u \sim l^{2/3}$. 

Since $\mathrm{d}x^2/\mathrm{d}t = 2\,\vec{x}(t)\cdot\vec{u}_\mathrm{L}(t) = 2\,\vec{u}_\mathrm{L}(t)\cdot\int_0^t \vec{u}_\mathrm{L}(t^\prime)\,\mathrm{d}t^\prime$, we can write $\mathrm{d}x_\mathrm{rms}^2/\mathrm{d}t = \mathrm{d} x^2/\mathrm{d}t = 2\,\int_0^t \vec{u}_\mathrm{L}(t)\cdot \vec{u}_\mathrm{L}(t-\tau_\mathrm{c})\,\mathrm{d}\tau_\mathrm{c}$, with $\tau_\mathrm{c} = t-t^\prime$. Now we perform an ensemble average over many particles and obtain $\mathrm{d}\langle x_\mathrm{rms}^2\rangle/\mathrm{d}t = 2\,\int_0^t \langle \vec{u}_\mathrm{L}(t)\cdot \vec{u}_\mathrm{L}(t - \tau_\mathrm{c}) \rangle \,\mathrm{d}\tau_\mathrm{c}$. The integrand is the Lagrangian velocity correlation tensor, and is a function of $\tau_\mathrm{c}$ only, if, as we have assumed, the turbulence is statistically steady, while the time-dependence is in the integration limit. The timescale over which the particle can be influenced by the flow turbulence is $t_\mathrm{L}$ and can be defined \citep[see][]{Davidson:2015} by extending the integration limit to infinity by $\langle u^2 \rangle \, t_\mathrm{L} = \int_0^\infty \langle \vec{u}_\mathrm{L}(t)\cdot \vec{u}_\mathrm{L}(t - \tau_\mathrm{c}) \rangle \,\mathrm{d}\tau_\mathrm{c}$. 
Now the transport of the dust particle depends on the relation of the timescale to $t_\mathrm{L}$, so that one can roughly distinguish two cases. 
If $t \ll t_\mathrm{L}$, the particle moves roughly ballistically since the correlation is strong and we have approximately 
$\langle \vec{u}_\mathrm{L}(t) \cdot\vec{u}_\mathrm{L}(t^\prime)\rangle \approx \langle u_\mathrm{L}^2\rangle \sim \langle u^2\rangle$ (using the appropriate eddy velocity fluctuation), with the correlation decaying exponentially as $t > t_\mathrm{L}$, resulting in $\langle \vec{u}_\mathrm{L}(t)\cdot \vec{u}_\mathrm{L}(t^\prime)\rangle \approx 0$. Therefore, $x_\mathrm{rms} \approx \sqrt{\langle u^2 \rangle} \,t$, for $t \ll t_L$, and for $t \gg t_L$, we can write $\mathrm{d}\langle x_\mathrm{rms}^2 \rangle/\mathrm{d}t  = 2\,\int_0^t \langle \vec{u}_\mathrm{L}(t)\cdot \vec{u}_\mathrm{L}(t - \tau_c) \rangle \,\mathrm{d}\tau_c \approx 2\, \int_0^\infty \langle \vec{u}_\mathrm{L}(t)\cdot \vec{u}_\mathrm{L}(t - \tau_\mathrm{c}) \rangle \,\mathrm{d}\tau_\mathrm{c}$, which yields  
$x_\mathrm{rms} \approx \sqrt{2\,\langle u^2\rangle\,t_\mathrm{L} \, t}$, reminiscent of random Markovian motion as it is proportional to $t^{1/2}$.

Applying this to the turbulent motion of dust particles in an expanding SB, we identify the integral scale of turbulence, $L$, with the SB radius $R(t)$ and the corresponding fluctuation velocity with $\sqrt{\langle u^2\rangle} \sim \dot{R}(t)$ for the largest energy-containing eddies. The average distance travelled by a particle then depends on the diffusion coefficient, which we estimate for homogeneous isotropic turbulence to be $\alpha_\mathrm{d} \sim q \, L \sqrt{\langle u^2 \rangle}$, where $q$ is a mixing constant of order unity ($0 \leq q \leq 1$). Consequently, the diffusion timescale is $\tau_\mathrm{d} \sim \alpha_\mathrm{d}/\langle u^2  \rangle$, and it turns out to be $\tau_\mathrm{d} \sim q \,L/\sqrt{\langle u^2  \rangle} = q \,R/\dot{R}$, which is approximately the dynamical expansion timescale of the LB. If, on the other hand, the dust particle crosses a large number of eddies, $\sqrt{\langle u^2\rangle}$ has to be replaced by the particle velocity $\sqrt{\langle u_\mathrm{L}^2\rangle}$, which can be significantly smaller.

As a final remark, we note that turbulent diffusion is very complex, and could be anomalous, for example, due to intermittency, leading to Lévy flight behaviour and superdiffusivity, in which the step size distribution is a tail-heavy power law, and where the square root of time in the expression for the RMS displacement has to be replaced by an exponent $\beta$ with $0.5 < \beta < 1$. This can be seen when examining the structure functions of order $p$, which characterise the type of turbulence, and are given by (for a simple illustration we assume incompressibility) $S_p(l) = \langle |\vec{u}(\vec{x} + \vec{l}) - \vec{u}(\vec{x})|^p \rangle$. In the inertial range, the structure functions are simply $S_p(l) \sim l^{\zeta(p)}$, with $\zeta(p) \sim p/3$, disregarding the fact that real turbulence is not space filling. We can then generalise our previous results for the two regimes. Within the correlation time $\tau$, the diffusion coefficient at scale $l$ is $\alpha_\mathrm{d}(l) \sim q \, l \sqrt{\langle u^2 \rangle} \sim l^{\zeta(2)} \, t$, as this is proportional to the structure function of second order, while for $t \gg \tau$, $\alpha_\mathrm{d}(l) \sim l \, l^{\zeta(1)} = l^{1+\zeta(1)}$. For Kolmogorov scaling, $\zeta \sim p/3$, we recover the well-known results for $\alpha_\mathrm{d}$. But now, we can allow for deviations from such a simple relation, for example, because higher-order structure functions may contribute significantly and non-locality becomes important, as the memory of previous eddy crossings determines the particle trajectories. Such an anomalous diffusion can be described by a fractional diffusion equation, which we will not discuss here further.

If the size of the dust particles becomes large, their transport by turbulent eddies, which is superdiffusive, becomes less efficient, so that it may become eventually subdiffusive. It should, however, be kept in mind that SN-generated dust grains are subject to heavy sputtering, and the surviving particles may be on average much smaller than those produced for example in the winds of asymptotic giant branch stars.

\section{Supplementary plots}
\label{app:B}

In this appendix, we present supplementary plots.

   \begin{figure*}
   \centering
   \resizebox{0.96\hsize}{!}
            {\includegraphics{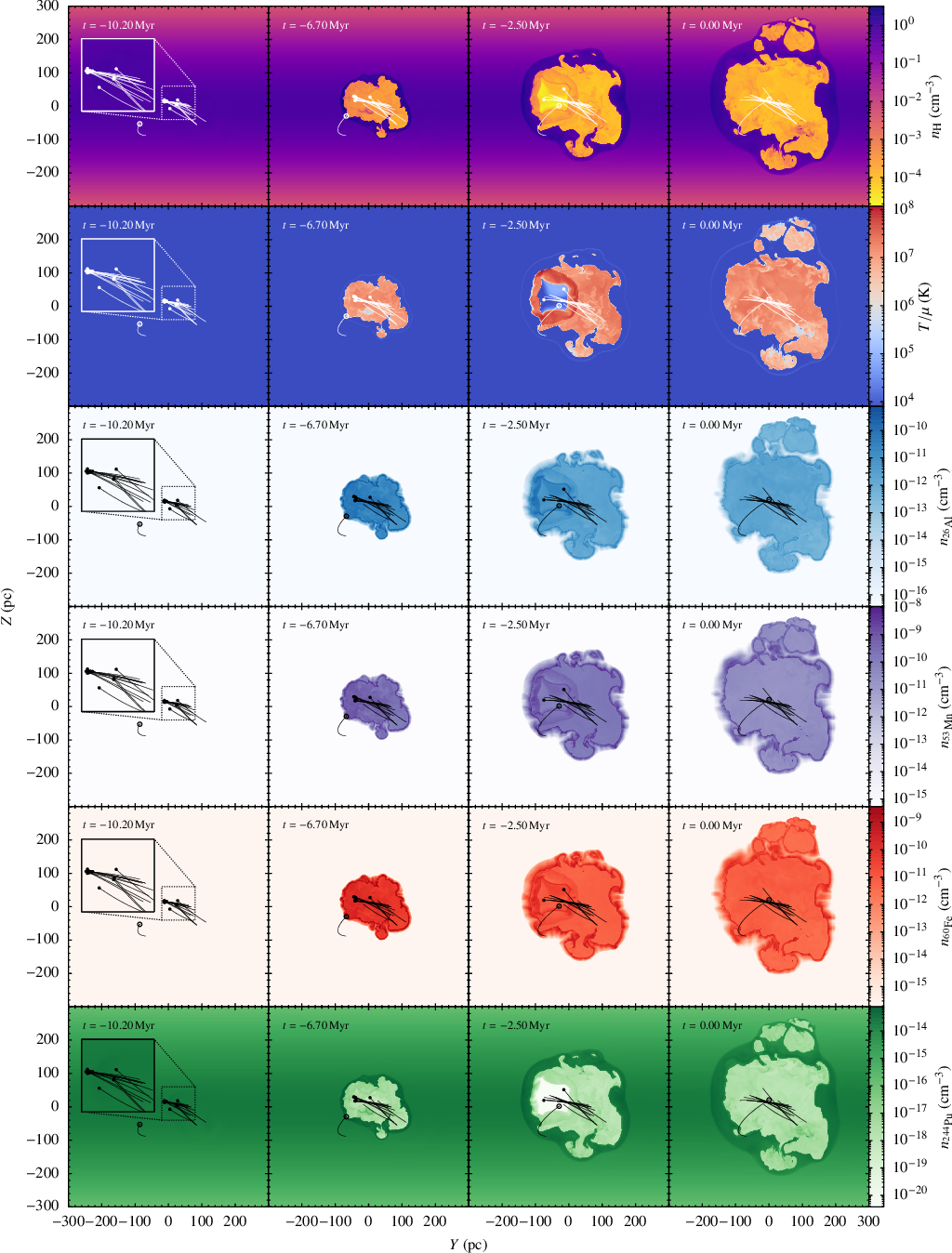}}
      \caption{As for Fig.~\ref{im:ts-sy}, but for slices at $X=0$.
      }
         \label{im:ts_sx}
   \end{figure*}

 \begin{figure*}
\sidecaption
  \includegraphics[width=12cm]{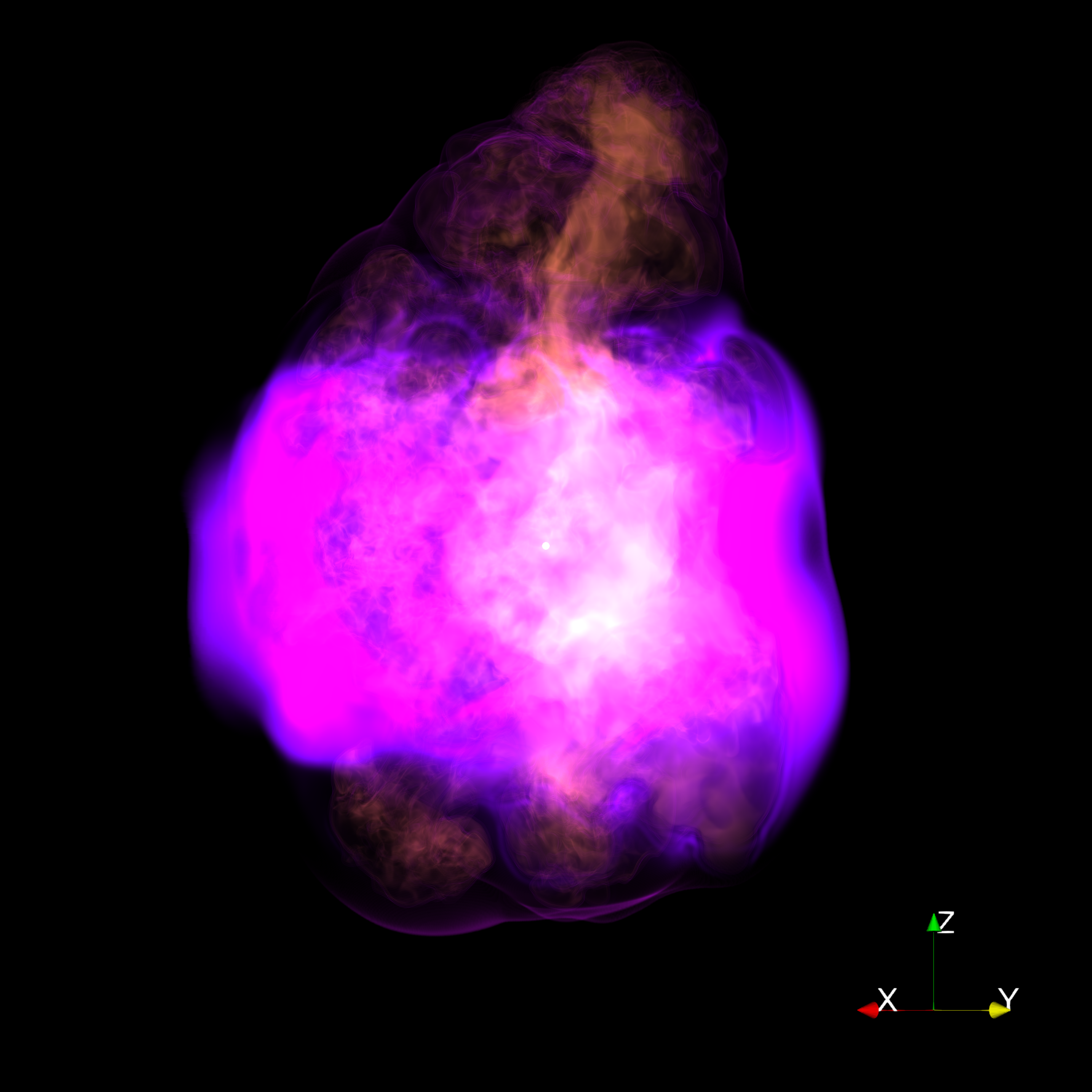}
     \caption{Volume rendering of the atomic hydrogen number density field associated with the LB (the quiescent inhomogeneous ambient medium is masked out) at $t=0$ (present time) in our fiducial simulation. The coordinate frame and colour coding is the same as in the corresponding slice plots (first row panels of Figs.~\ref{im:ts-sy}, \ref{im:ts-sz}, and \ref{im:ts_sx}). The white dot in the centre of the image marks the position of the Sun.}
     \label{im:nhrendering}
\end{figure*}

   \begin{figure*}
   \centering
   \resizebox{\hsize}{!}
            {\includegraphics{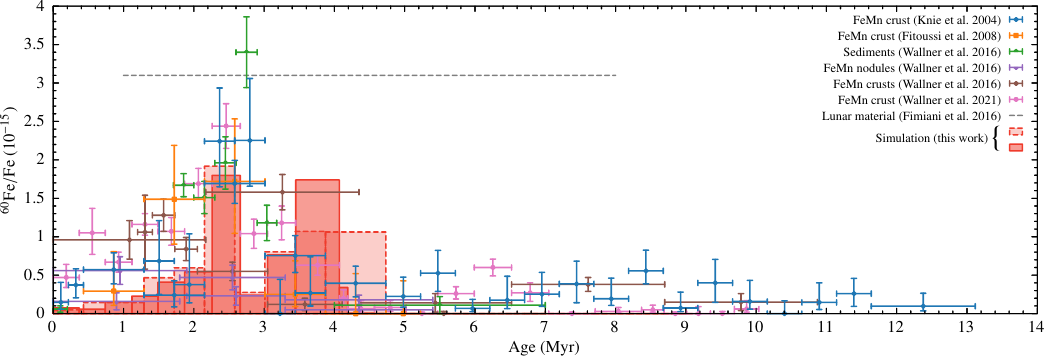}}
      \caption{As for Fig.~\ref{im:crustplot}a, except that the simulation was carried out without the LB progenitor star with the lowest initial mass ($\unit[12.85]{M_\sun}$), so that the last SN explosion occurred $\unit[1.68]{Myr}$ ago instead of $\unit[0.88]{Myr}$ ago.}
         \label{im:crustplotapp}
   \end{figure*}

\end{appendix}
\end{document}